\documentclass[twocolumn]{aastex631}

\usepackage{inputenc}
\usepackage{booktabs}
\usepackage{color, amssymb, amsmath, natbib, color} 
\usepackage{multirow, placeins, subfigure}
\usepackage{graphicx}
\usepackage{epstopdf}
\usepackage{bm}
\usepackage{overpic}
\usepackage{hyperref}
\usepackage{comment}

\definecolor{lgray}{gray}{0.3}
\definecolor{orange}{rgb}{1,0.5,0}

\def\degree{{^\circ}}
\newdimen\digitwidth
\setbox1=\hbox{0}
\digitwidth=\wd1
\catcode`"=\active
\def"{\kern\digitwidth}

\newcommand\xone{x_1}
\newcommand\xtwo{x_2}
\newcommand\xthree{x_3}

\newcommand\vl{v_{\rm{LOS}}}

\newcommand\Msun{\; {M}_{\odot}}
\newcommand\Rsun{{R}_{0}}

\newcommand\kms{\; {\rm km}\;{\rm s}^{-1}}

\newcommand\pc{\;{\rm pc}}

\newcommand\kpc{\;{\rm kpc}}

\newcommand\freq{\kms\kpc^{-1}}

\newcommand\Gyr{\;{\rm Gyr}}

\newcommand\C{{\mathcal C}}

\newcommand\simgt{\lower.5ex\hbox{$\; \buildrel > \over \sim \;$}}
\newcommand\simlt{\lower.5ex\hbox{$\; \buildrel < \over \sim \;$}}
\newcommand{\RNum}[1]{\uppercase\expandafter{\romannumeral #1\relax}}

\newcommand\lv{{l-v}}

\defcitealias{P17}{P17}
\defcitealias{sormani_new}{H24}
\defcitealias{sanders2023}{S23}
\defcitealias{shen10}{S10t800}

\begin{document}


\title{The Milky Way bar potential constrained by the kinematics of SiO maser stars in BAaDE Survey}


\author[0000-0003-1309-3050]{Tian-Ye Xia}
\affiliation{Department of Astronomy, School of Physics and Astronomy, Shanghai Jiao Tong University\\ 
800 Dongchuan Road, Shanghai 200240, China}
\affiliation{Key Laboratory for Particle Astrophysics and Cosmology (MOE) / Shanghai Key Laboratory for Particle Physics and Cosmology\\
Shanghai 200240, China}


\author[0000-0001-5604-1643]{Juntai Shen}
\affiliation{Department of Astronomy, School of Physics and Astronomy, Shanghai Jiao Tong University\\ 
800 Dongchuan Road, Shanghai 200240, China}
\affiliation{Key Laboratory for Particle Astrophysics and Cosmology (MOE) / Shanghai Key Laboratory for Particle Physics and Cosmology\\
Shanghai 200240, China}
\affiliation{Shanghai Astronomical Observatory, Chinese Academy of Sciences, 80 Nandan Road, Shanghai 200030, China}
\email{jtshen@sjtu.edu.cn}

\author[0000-0002-0627-8009]{Zhi Li}
\affiliation{Shanghai Key Lab for Astrophysics, Shanghai Normal University, 100 Guilin Road, Shanghai 200234, People's Republic of China}
\affiliation{Department of Astronomy, School of Physics and Astronomy, Shanghai Jiao Tong University, 800 Dongchuan Road, Shanghai 200240, People's Republic of China}
\email{lizhi@shnu.edu.cn}

\author[0009-0001-4509-0025]{Huaijin Feng}
\affiliation{Department of Astronomy, School of Physics and Astronomy, Shanghai Jiao Tong University\\ 
800 Dongchuan Road, Shanghai 200240, China}
\affiliation{Key Laboratory for Particle Astrophysics and Cosmology (MOE) / Shanghai Key Laboratory for Particle Physics and Cosmology\\
Shanghai 200240, China}

\author[0000-0003-3096-3062]{Lor\'ant O.\ Sjouwerman}
\altaffiliation{Adjunct Professor at the Department of Physics and
  Astronomy, the University of New Mexico}
\affiliation{National Radio Astronomy Observatory, Pete V.\ Domenici
  Science Operations Center Operations Center, Socorro, NM 87801, USA}

\author[0000-0003-0615-1785]{Ylva M.\ Pihlstr\"om}
\affiliation{Department of Physics and Astronomy, the University of
  New Mexico, Albuquerque, NM 87131, USA}
\affiliation{National Radio Astronomy Observatory, Pete V.\ Domenici
  Science Operations Center Operations Center, Socorro, NM 87801, USA}

\author[0000-0002-8069-8060]{Megan O.\ Lewis}
\altaffiliation{Recently moved from Nicolaus Copernicus Astronomical
  Center, Polish Academy of Sciences, Warszawa, Poland}
\affiliation{Leiden University, Leiden Observatory, Leiden, 2300RA,
  The Netherlands}

\author[0000-0002-3019-4577]{Michael C.\ Stroh}
\affiliation{Northwestern University, Center for Interdisciplinary
  Exploration and Research in Astrophysics, Evanston, IL 60201, USA}

\begin{abstract}

We introduce a novel method that utilizes the longitude-velocity ($\lv$) envelope to constrain the Milky Way (MW) bar potential. Previous work \citep{Habing2016} used the $\lv$ diagram to explain the distribution of the observed high-velocity stars. We successfully reproduce their results, but find that their method is limited to only one single type of periodic orbits. In contrast, we propose that the $\lv$ envelope provides much more comprehensive constraints. We compare the properties of test particles in the \citet{P17} MW potential model \citepalias{P17} with the observed SiO maser stars from the Bulge Asymmetries and Dynamical Evolution (BAaDE) survey. We find that the $\lv$ envelope generated by the bar potential demonstrates reasonable agreement with the observational data, albeit with slight discrepancies near the Galactic center. The inconsistencies suggest that the P17 potential yields a lower central rotation curve, a slightly larger quadrupole strength, or a possibly underestimated pattern speed. {We also adopt an updated version of the \citetalias{P17} potential with a modified central mass component (CMC) proposed by \citet{sormani_new} (\citetalias{sormani_new}). The fitting of the $\lv$ envelope suggests that the \citetalias{sormani_new} potential does not completely address the existing challenges and may hint at a possible underestimation of the central bar mass.} Our study demonstrates that the $\lv$ envelope can be used as a valuable tool for constraining the Galactic potential and provides insights into the Milky Way bar potential.

\end{abstract}


\section{Introduction}

The existence of the Galactic bar was confirmed in the 1990s \citep{Blitz1991, Binney1991}. Since then, significant progress has been made in observational data and modeling of stars and gas, greatly improving our understanding of the Galactic bar \citep{biney1997, Launhardt2002,shen10, Kunder_2012,Sormani2015, Li_2016, P17, bovy2019, shenzheng2020, Li_2022, gaiadr3}. Despite these achievements, the gravitational potential of the Milky Way bar remains less understood, posing a challenge for theoretical investigations into the dynamics of stars and gas. 

There have been a number of studies aiming to constrain the Galactic bar potential. From the gas side, \citet{Sormani2015} demonstrated that the quadrupole term of the Galactic bar potential can be constrained by analyzing the gas flow in the $\lv$ diagram, which shows the gas intensity as a function of Galactic longitude ($l$) and line-of-sight velocity ($\vl$) \footnote{In this paper, the term "radial velocity" refers to a model parameter representing the velocity component relative to the dynamical center of the Galaxy. On the other hand, ``line-of-sight velocity" is used to describe the velocity component relative to the observer, typically measured through the spectral Doppler shift.}. To achieve this, they employed a bar potential consisting of only the monopole and quadrupole terms. They varied the quadrupole term while keeping the monopole term unchanged. Their investigation yielded well-constrained values for the scale length ($1.5 \kpc$) and pattern speed ($40 \freq$) of the bar. 

From the stellar side, \citet{Habing2016} utilized closed elliptical orbits in a homogeneous ellipsoidal bar, commonly referred to as the Freeman bar \citep{freeman1, freeman2}, to account for the observed high-velocity ($\vl \gtrsim 200\kms$) stars in the $\lv$ space at the Galactic center. The model proposed by \citet{Habing2016} provided a straightforward and quantitative explanation for the high-velocity stars, as well as constraints regarding the eccentricity or average stellar density in the bar. These high-velocity stars may also be related to the high-velocity peaks in the bulge that are discussed by \citet{MM2015, AS2015, Zhou_2017} through $N$-body simulations and observation analyses.

Nevertheless, there are two main limitations in \citet{Habing2016}: (1) Their model aimed to account for the high-velocity streams relied only on closed elliptical orbits, whereas the actual orbits in a Freeman bar are the superposition of two ellipses \citep{freeman1, freeman2}. Upon adopting the best-fit parameters recommended by \citet{Habing2016} into the Freeman bar model, we find that certain orbits, apart from the closed elliptical ones, may generate stars with extremely high LOS velocities well above $400\kms$, which have never been observed. (2) When we consider the distribution of all orbits in the $\lv$ diagram, we find that the overall shape of the $\lv$ envelope generated by the Freeman bar significantly deviates from the observation results. The details can be found in \S~\ref{section:Freeman}.

Although \citet{Habing2016} restricted the fitting comparison to the idealized Freeman bar potential, his work still gives us some insight into constraining the Galactic potential in the $\lv$ diagram. We have noticed that orbits generated by the bar potential tend to exhibit a relatively well-defined $\lv$ envelope when all types of orbital families are considered. This motivates us to use the shape of the $\lv$ envelope to constrain the bar potential.


In addition to the aforementioned studies, there have been other investigations that have presented more intricate and realistic models of the Milky Way bar based on observational stellar data. Notably, the work conducted by \citet{P17} (hereafter \citetalias{P17}) constructed dynamical models for the Galactic bulge, bar, and inner disc using the made-to-measure (M2M) method. Their models were specifically designed to accurately reproduce the density distribution and kinematics of red giant clumps in the Galactic bulge. \citet{Sormani2022a} built upon this work by introducing an analytical model of the Galactic bar, which relied on an analytical approximation of the density distribution in \citetalias{P17}. They implemented the analytical form of the \citetalias{P17} potential into the python package {\tt Agama} \citep{agama_2019}.  

The M2M model developed by \citetalias{P17} well predicted the stellar kinematics in the bulge and bar region. However, it exhibited poor agreement with the gas dynamics in the central region, as highlighted by \citet{Li_2022}. This discrepancy arises from the fact that \citetalias{P17} employed an elongated exponential disk as the nuclear component, whereas observations and simulations indicate that gas models tend to favor a nearly axisymmetric central mass distribution \citep{Henshaw2016, Li_2022}. \citet{Li_2022} replaced the central mass component (CMC) in \citetalias{P17} with the fiducial model of \citet{Sormani_2020}, which is composed of the Nuclear Stellar Cluster (NSC) from \citet{nsc2015} and the Nuclear Stellar Disk (NSD) derived by Jeans modeling. \citet{sormani_new} applied this modification to the analytical \citetalias{P17} potential, resulting in a new potential (hereafter \citetalias{sormani_new}). The \citetalias{sormani_new} potential is also included in {\tt Agama}. In our investigation, we adopt both the \citetalias{P17} and \citetalias{sormani_new} potentials as inputs.

We use the observational data from the Bulge Asymmetries and Dynamical Evolution (BAaDE) survey \citep{2015Sjouwerman, 2017Sjouwerman, trapp_2018, Stroh_2019, Lewis2021}. This dataset comprises information on $\sim 15,000$ SiO maser stars \citep{Lewis2021}. SiO maser stars are excellent tracers of the Galactic bulge as they do not suffer severely from dust extinction and can provide precise LOS velocities. 


Our goal in this paper is to develop a novel method based on the $\lv$ diagram to constrain the Milky Way bar potential. We demonstrate that the $\lv$ envelope can be a promising constraint after investigating the $\lv$ diagram of actual orbital families in a Freeman bar. To verify its effectiveness, we compare the $\lv$ envelope derived from the \citetalias{P17} potential, employing carefully selected initial conditions, with the observational data. Our findings reveal a generally good agreement between the $\lv$ envelope of the \citetalias{P17} potential and the observed data, with only minor discrepancies. We further investigate the factors influencing the envelope and draw implications about the \citetalias{P17} potential. Additionally, we explore the \citetalias{sormani_new} potential and find that it does not entirely resolve the discrepancies highlighted by the envelope comparison. Our investigation suggests that incorporating an extra elongated central massive component (CMC) could potentially help alleviate the problem. Overall, our results robustly confirm the utility of the $\lv$ envelope to constrain the bar potential of the Milky Way.

The paper is organized as follows: we provide details of the maser data in \S~\ref{section:data}. We examine the $\lv$ diagram of the Freeman bar in \S~~\ref{section:Freeman}. We present the comparison between the $\lv$ envelope derived from the \citetalias{P17} potential and the observational data in \S~~\ref{section:comparison}. We explore the factors influencing the $\lv$ envelope in \S~~\ref{section:factor}. {We discuss the implication of the $\lv$ envelope in constraining the Galactic bar potential in \S~\ref{section:modification}.} We summarize in \S~~\ref{section:conclusion}.

\section{Observational results of maser stars}
\label{section:data}

\subsection{l-v diagrams of maser stars}
\label{section:lv_define}

We use the observational data of SiO maser stars from the Bulge Asymmetries and Dynamical Evolution (BAaDE) survey \citep{2015Sjouwerman, trapp_2018,Stroh_2019,Lewis2021,Sjouwerman2024}. Masers are luminous radio beacons found in the advanced stages of Asymptotic Giant Branch (AGB) stars. Compared with the traditional OH masers, SiO masers are more prevalent due to their occurrence in stars with varied mass loss rates, lower initial mass and higher metallicity. Furthermore, SiO masers serve as excellent tracers of the Galactic bulge, as they are less affected by external environments such as areas with significant extinction and provide precise LOS velocities.

The source list used in the BAaDE survey was derived from an infrared catalog (MSX), which introduces sensitivity to the MSX observational constraints. The far side data is more susceptible to extinction effects, and the challenge of identifying individual point sources due to limited sensitivity and a large beam size. In addition, the lower metallicity outside the solar radius makes the SiO masers less likely to be found. Thus, the maser data exhibit a sampling bias, with inadequate coverage on the far side of the bulge regions compared to the near side, as noted by \citet{maser_distance_2024}.

\begin{figure}[htbp!]
  \centering
  \includegraphics[width=0.44\textwidth]{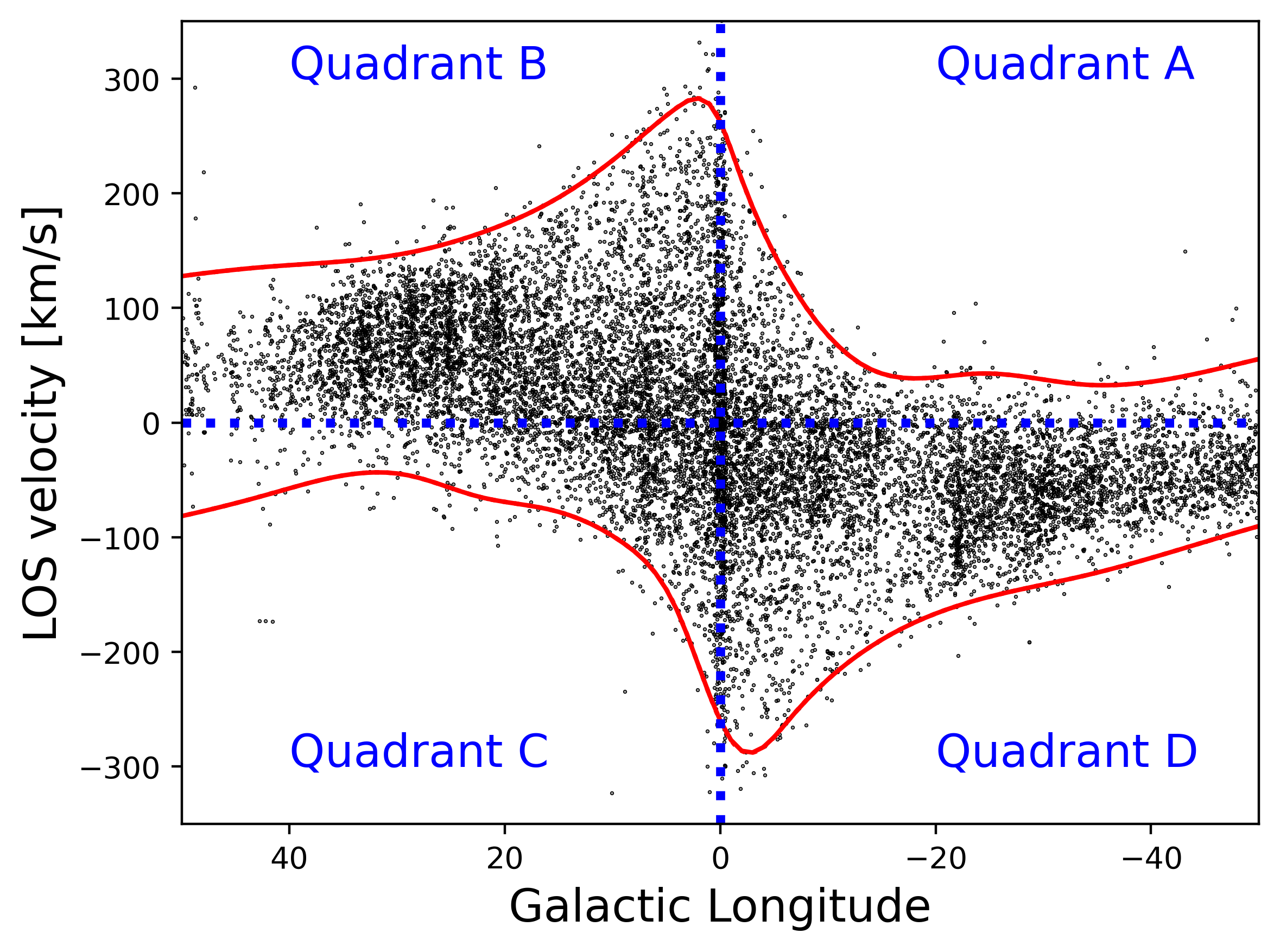}
  \includegraphics[width=0.44\textwidth]{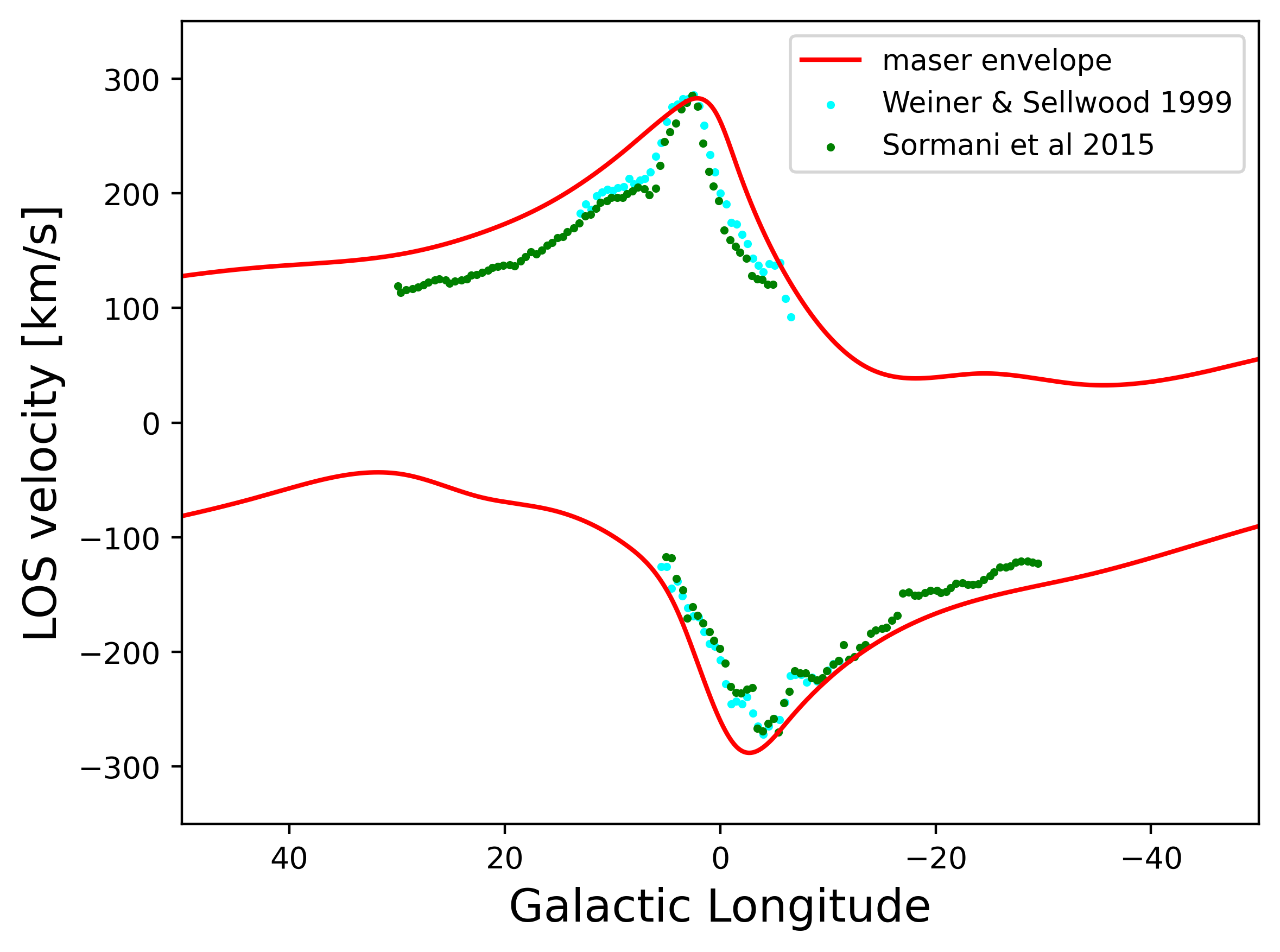}\\

  \caption{The $\lv$ envelope of the SiO maser stars from the BAaDE survey and several gas observations. The red curves in the two panels are maser envelopes defined in \S~\ref{section:lv_define}. The upper panel shows the $\lv$ diagram of the SiO maser stars, where the black points represent the data points from the BAaDE survey. The blue dashed lines divide the $\lv$ diagram into four regions, defined as Quadrants A, B, C and D, respectively. The lower panel shows the comparison between the maser and gas envelopes. {The cyan and green points are HI envelopes manually extracted from \citet{WS99} and \citet{sormani2015gas}, respectively.}  }
  \label{fig:maser_lv}
\end{figure}

The BAaDE survey has provided us 15,207 SiO maser stars with their preliminary LOS velocities ($\vl$), Galactic longitudes ($l$), and latitudes ($b$). Figure~\ref{fig:maser_lv} plots the $\lv$ distribution of the SiO masers, which exhibits a well-defined envelope represented by the red curve.


We define the region of $l<0\degree$ and $\vl>0\kms$ as Quadrant A, the region of $l>0\degree$ and $\vl>0\kms$ as Quadrant B, the region of $l>0\degree$ and $\vl<0\kms$ as Quadrant C, and the region of $l<0\degree$ and $\vl<0\kms$ as Quadrant D.

Notably, the number of masers at $20\degree<l<40\degree$ is approximately $20\%$ larger than that at $-40\degree<l<-20\degree$, which introduces a slight asymmetry in the envelope. We further discuss this in Appendix~\ref{section: asymmetry}. 

To define the $\lv$ envelope, we employ the quantile regression, which is a statistical method used to estimate the conditional quantiles of the response variables, rather than the mean value as the least squares regression does \citep{Quantile_Regression}. This approach is commonly used as an extension of linear regression when the assumptions and limitations of linear regression cannot fully meet the requirements of the analysis. In this study we use the python package 'statsmodels' to derive the quantiles of the $\lv$ distribution \citep{statsmodels}. Specifically, we set the upper and lower limits to correspond to the $99\%$ and $1\%$ quantiles, respectively, as the $\lv$ envelope. The degree of freedom for the fit is chosen to be $16$, which helps to accurately capture the shape and variability of the envelope. We have also explored alternative choices for the degree of freedom, but found them either resulting in loss of information or excessive noise in the envelope fitting.

\subsection{Maser observation compared with gas surveys}

A natural question is how stars and gas are distributed differently in the $\lv$ space. We conduct a comparison between the SiO maser envelope and the gas envelopes from several gas surveys. The gas data consist of HI $\lv$ envelope extracted from \citet{WS99} and \citet{sormani2015gas}.


The maser envelope appears to be slightly larger than the gas envelopes. This discrepancy can be attributed to the different velocity dispersions of the two tracers. SiO maser stars encompass a range of ages from $0.2$ to $2\Gyr$, including more evolved stars that may exhibit larger velocity dispersions. In contrast, gas typically has lower temperatures and smaller velocity dispersions, which can result in a narrower $\lv$ envelope (see \S~\ref{section:sigma} for a detailed discussion).




\section{$\lv$ diagram of the Freeman bar}
\label{section:Freeman}

The Freeman bar \citep{freeman1,freeman2} is the simplest analytical bar, which is essentially the same with the homogeneous ellipsoidal utilized in \citet{Habing2016}. It generates a quadratic potential given by \citet{freeman2}:
\begin{equation}
 \Phi(x,y)=\frac{1}{2}(\Omega_x^{2}x^{2}+\Omega_y^{2}y^{2}),
\end{equation}
where $\Omega_x$ and $\Omega_y$ is related to the eccentricity $e$ and density $\rho$ of the bar. The potential is rotating with a fixed pattern speed $\Omega_{\rm{b}}$.

A general orbit in a Freeman bar is the superposition of two ellipses (i.e.\ 2 orbital families): the prograde $\alpha-$ellipse along the bar axis (or $x_1$ orbits) and the retrograde $\beta-$ellipse perpendicular to the bar axis (or $x_4$ orbits). There is no $x_2$ or $x_3$ orbit in the Freeman bar.

In order to explain the observed high-velocity stars, \citet{Habing2016} utilized only the closed $\alpha-$ellipse orbits in their analysis of the $\lv$ diagram. They explored different values of eccentricity and density for the Freeman bar potential, assessing whether the $\lv$ diagram of the closed $\alpha-$ellipse orbit passing through a fixed point in the phase space could encompass all the observed high-velocity stars. They proposed the best-fit parameters for the bar potential as $e=0.99$ and $\rho=15\Msun\pc^{-3}$.

We adopt exactly same parameters as used in \citet{Habing2016} when plotting the $\lv$ diagram of the best-fit Freeman bar: the pattern speed is $73\freq$, the rotation velocity of the Sun is $255\kms$ and the solar radius ($\Rsun$) is $8.34\kpc$. the Sun is assumed to be in a perfect circular motion. The angle between the major axis of the bar and the Sun-Galactic center line is set to be $25\degree$. The unit of length is $\kpc$, and the unit of velocity is $\kms$.



To some extent, \cite{Habing2016} provided an explanation for several specific high-velocity stars, offering insights into their origins. However, their methods have two main limitations.

The first limitation is that they restricted their test orbits to only one periodic orbital family, i.e.\ the $\alpha-$ellipses ($x_1$ orbits), neglecting the presence of the $\beta-$ellipses ($x_4$ orbits) in the Freeman bar potential. Thus, we extend the findings of \citet{Habing2016} to encompass all orbital families of the Freeman bar. 

\begin{figure*}[htbp!]
  \centering
  \includegraphics[width=0.88\textwidth]{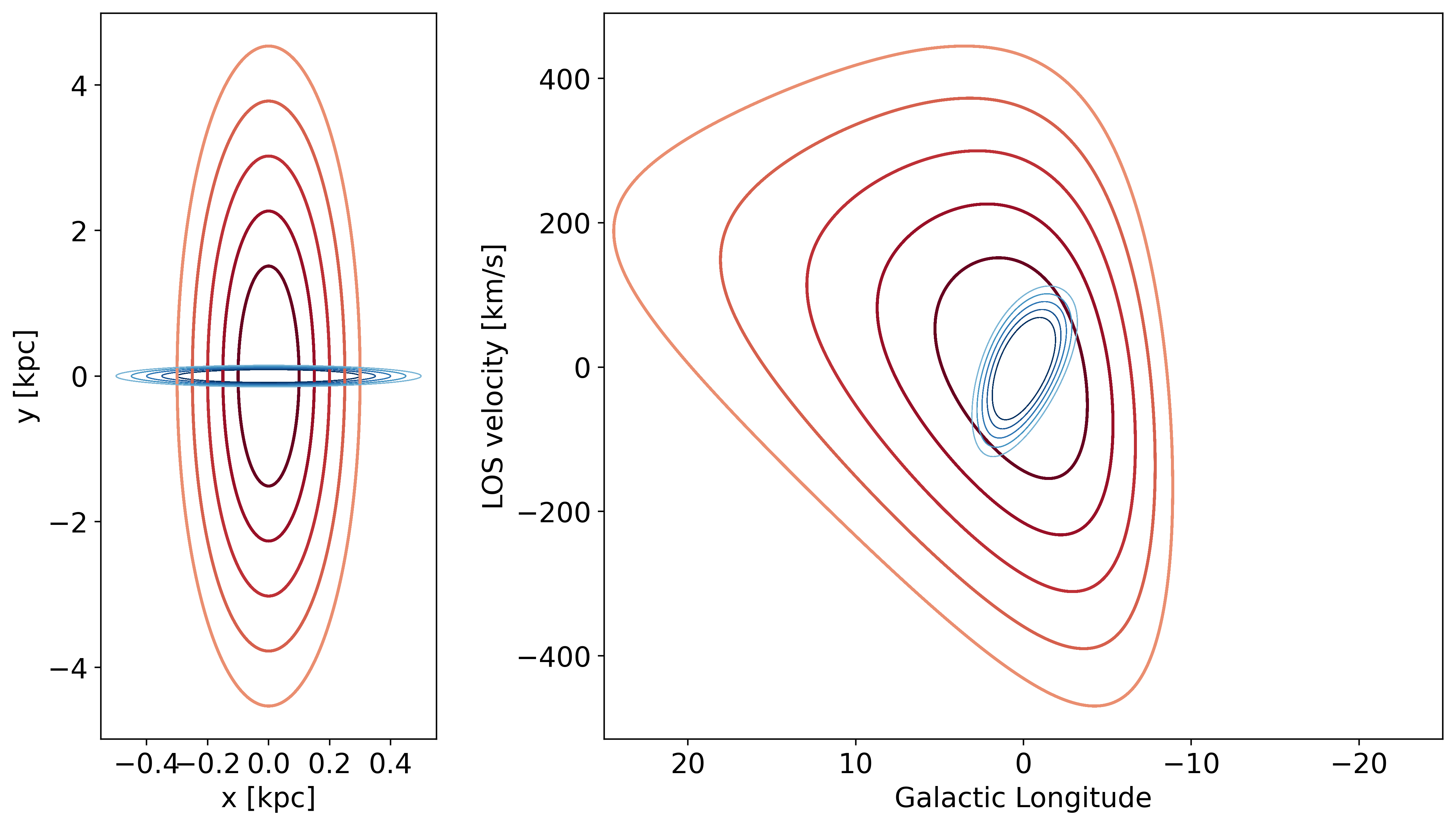}
  \caption{Periodic orbits in the Freeman bar and the corresponding $\lv$ diagram. The parameters used are the same as in \citet{Habing2016}. The bar major axis is along the $y$-axis. The left panel shows the $\alpha-$ellipses along the bar axis ($x_1$ orbits) denoted by red curves and the $\beta-$ellipses perpendicular to the bar axis ($x_4$ orbits) presented by blue curves. The right panel shows the corresponding $\lv$ diagram of the orbits. The $\alpha-$ellipses and the $\beta-$ellipses exhibit two loop families with different directions of extension.  }
  \label{fig:freeman_lv}
\end{figure*}

We first consider the periodic orbits, which are limited to the closed $x_1$ and $x_4$ orbits in the Freeman bar. These orbits give rise to two distinct loop families in the $\lv$ diagram, as depicted in Figure~\ref{fig:freeman_lv}. We can indeed reproduce the results of \citet{Habing2016}; the red $\lv$ contours in Figure~\ref{fig:freeman_lv} match Figure 6 of \citet{Habing2016} nicely. Notably, we observe that the largest $x_1$ orbit displayed in the right panel of Figure~\ref{fig:freeman_lv} can exhibit extremely high LOS velocities greater than $400\kms$.


\begin{figure*}[htbp!]
  \centering
  \includegraphics[width=0.44\textwidth]{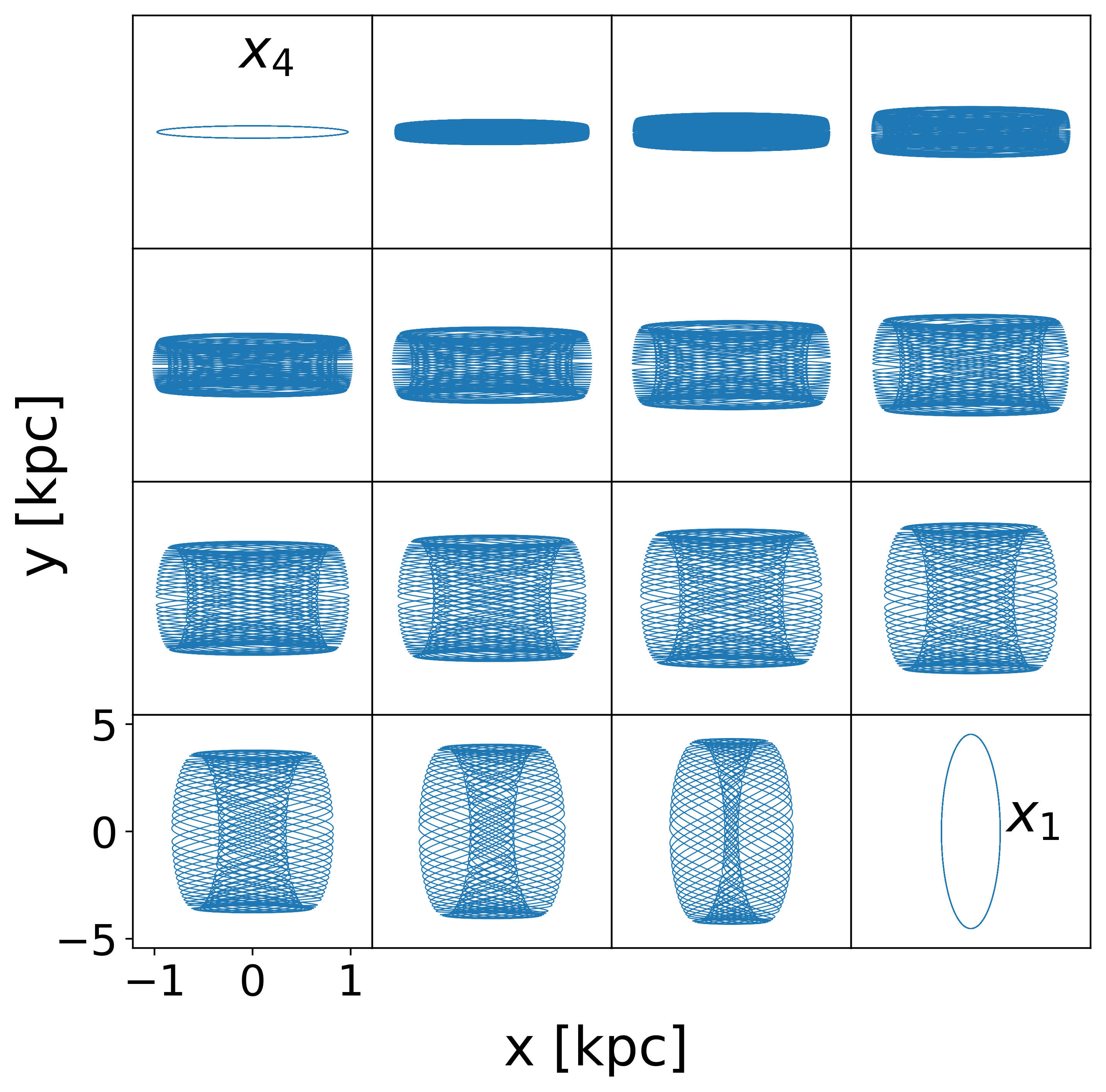}
  \includegraphics[width=0.44\textwidth]{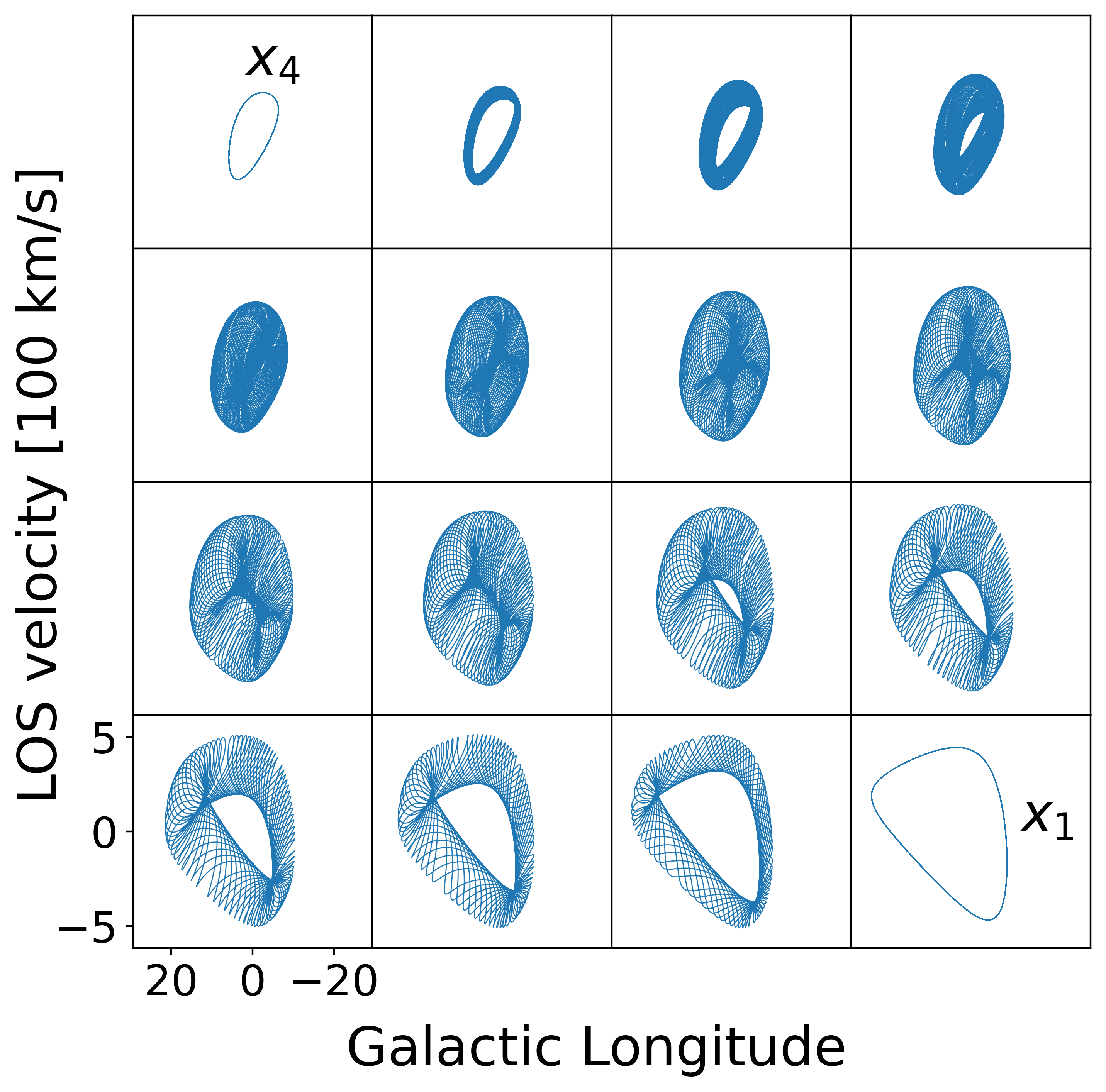}\\

  \caption{Orbit evolution in the Freeman bar and the corresponding $\lv$ diagram. The parameters used in model are the same with \citet{Habing2016}. The bar major axis is along the $y-$axis. The left panel, from upper left to lower right, shows the orbit evolution trend from $x_4$ (top left) to $x_1$ (lower right) orbit under a given Jacobi energy. The right panel shows the corresponding evolution in the $\lv$ diagram.}
  \label{fig:freeman_evolve_lv}
\end{figure*}

Next, we shift our focus to the general orbits within the Freeman bar and examine how their orbital structure is distributed in the $\lv$ diagram. The results are presented in Figure~\ref{fig:freeman_evolve_lv}. We find that the $\lv$ diagram of a general orbit in the Freeman bar is the superposition of the loop generated by the $\alpha-$ellipse and the loop generated by the $\beta-$ellipse. As the orbit elongation changes from perpendicular to parallel to the bar major axis, we note that the highest LOS velocity can reach approximately $500\kms$.

It is worth noting that the Freeman bar has a linearly rising rotation curve. The best-fit potential yields an unrealistically high rotation velocity at solar radius ($\sim 4000\kms$). This leads to the extremely high LOS velocities, as shown in Figure~\ref{fig:freeman_lv} and \ref{fig:freeman_evolve_lv}.

\begin{figure}[htb]
  \centering
  \includegraphics[width=0.44\textwidth]{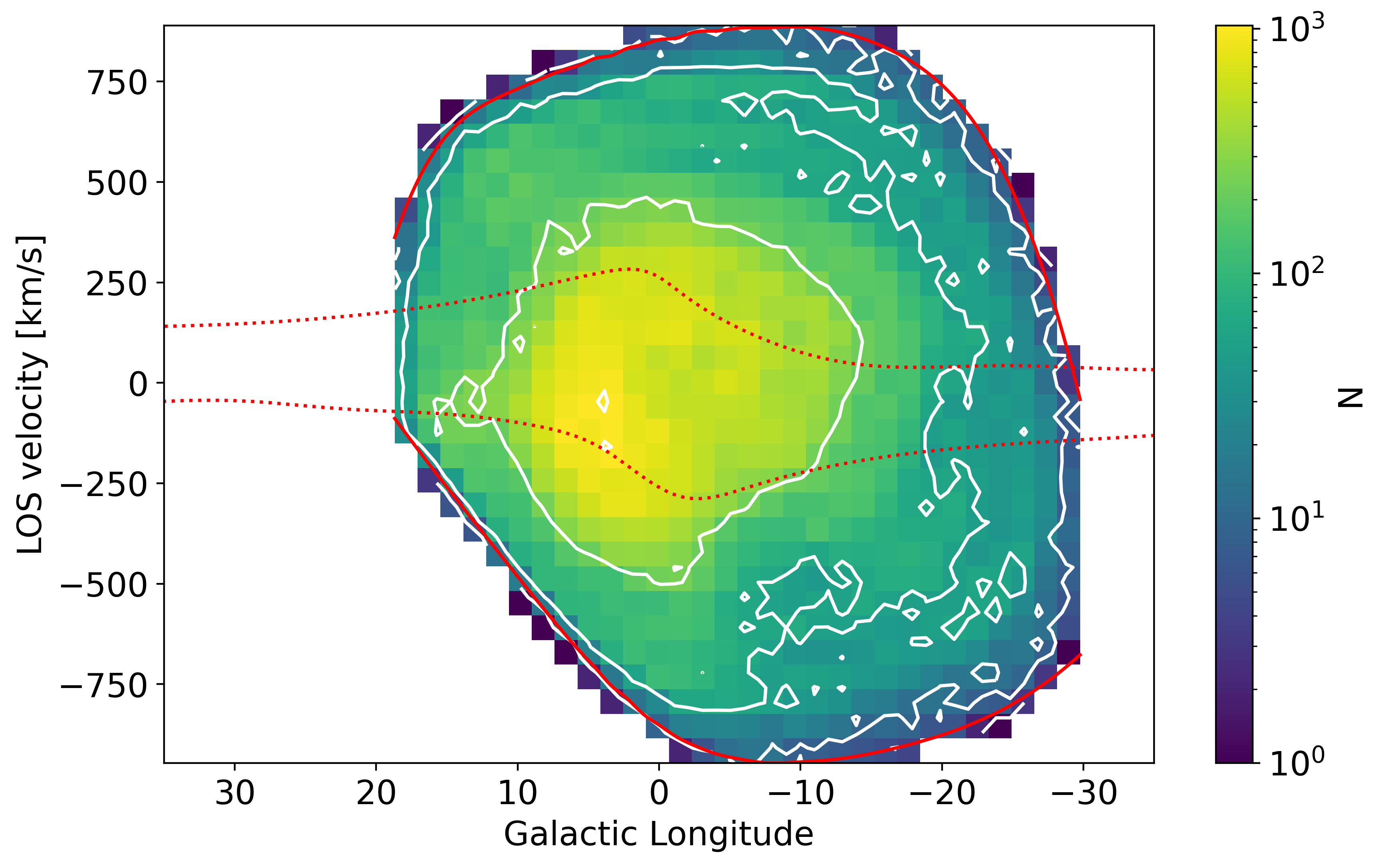}\\

  \caption{The $\lv$ diagram of orbits in the Freeman bar model. Only orbits inside $0.5\Rsun$ are included in the $\lv$ diagram. The color represents the number density. The white curve shows the contour of the number density. The red solid curve represents the $\lv$ envelope of the model, and the red dashed line represents the maser envelope. }
  \label{fig:freeman_tot_lv}
\end{figure}

The second limitation is that the $\lv$ envelope generated by the Freeman bar differs significantly from the observed maser envelope, as shown in Figure~\ref{fig:freeman_tot_lv}. The orbits are produced by randomly assigning initial conditions following a Gaussian distribution. The distribution is centered on $R=3\kpc$ with $\sigma_r=1\kpc$, $v_r=0\kms$ with $\sigma_{v_r}=40\kms$ and $v_\phi=0\kms$ with $\sigma_{v_\phi}=40\kms$, defined in the co-rotating frame. Only orbits within $0.5\Rsun$ are included in the $\lv$ diagram. The overall distribution in the $\lv$ diagram exhibits a roundish envelope with unrealistically high maximum velocities, which is significantly different from the observed envelope (the red line in Fig.~\ref{fig:maser_lv}).

Our analysis reveals that different orbital families in a Freeman bar can exhibit significantly different morphologies in the $\lv$ diagram. Although the periodic $x_4$ orbits in Figure~\ref{fig:freeman_lv} may not influence the envelope for its small size, the non-periodic (general) orbits in Figure~\ref{fig:freeman_evolve_lv} affect significantly the morphology of the envelope. 

The idealized nature of the Freeman bar limits the usefulness of its $\lv$ envelope for providing meaningful constraints. Therefore, it is more interesting to employ a relatively well-constrained Milky Way bar model to investigate how sensitive the $\lv$ envelope depends on the bar potential.



\section{$\lv$ envelope of the \citetalias{P17} potential compared with data}
\label{section:comparison}

In order to obtain a more realistic comparison with the observational data, we employ the \citetalias{P17} potential proposed by \citet{P17} and parametrized by \citet{Sormani2022a}.

The \citetalias{P17} potential provides an analytical fit to the density distribution in the \citetalias{P17} model, offering a comprehensive representation of the Galactic bar structure. It presents a three dimensional density which includes three bar components as well as an axisymmetric disk \citep{Sormani2022a}. By utilizing this potential, we aim to explore the properties of the envelope in a more accurate and representative manner. 

The \citetalias{P17} potential is available in the python package {\tt Agama} \citep{agama_2019}, which we use to perform the orbital integration.

Following \citetalias{P17}, {we set the pattern speed of our standard model to be $37.5\freq$} and the angle between the major axis of the bar and the Sun-Galactic center line to be $28\degree$. We assume that the rotation velocity of the Local Standard of Rest (LSR) is $238\kms$ and the solar radius is $8.15\kpc$. The vertical height of the Sun is $0.025\kpc$. The solar motions relative to the LSR in tangential, radial and vertical directions are $10.6\kms$, $10.7\kms$, $7.6\kms$, respectively. These values are taken from \citet{Schonrich2010, Bland-Hawthorn2016, Reid2019, gravity2019}. The unit of length is $\kpc$, and the unit of velocity is $\kms$. Note that the parameters used here are different from \S~\ref{section:Freeman}, where we adopt the same parameters utilized in \citet{Habing2016}.

\subsection{Setting of initial conditions}
\label{section: initialcondition}

Accurate reproduction of the observational results in our simulations requires careful consideration of the initial conditions, as they have a significant impact on the resulting $\lv$ envelope. To achieve this, we ensure that the distributions of the test particles integrated from our initial conditions closely resemble the velocity dispersion distributions provided by the Gaia Data Release 3 (Gaia DR3) data \citep{gaiadr3}.  

{The initial conditions are set as follows: \romannumeral1) The initial cylindrical radius $R$ follows the stellar radial surface density profile of the \citetalias{P17} potential. The initial azimuthal angle $\phi$ is randomly distributed between $0$ and $2\pi$. The initial vertical height $z$ is sampled from Gaussian distributions with a mean value of $\overline{z}=0 \kpc$ and a standard deviation of $\sigma_z=0.3 \kpc$. \romannumeral2) The initial velocities in all directions are determined by different Gaussian distributions, with the dispersion varying as a function of radius. Specifically, the radial velocity has a mean value of $\overline{v}_R=0 \kms$ and a standard deviation of $\sigma_{v_R}(R)=60\kms\times\exp{(-R/12\kpc)}$. The azimuthal velocity has a mean value of $\overline{v}_{\phi}=V_c(R)-{\sigma_{v_R}}^2/80\kms$, where $V_c$ is the rotation velocity of the potential, and a standard deviation of $\sigma_{v_\phi}=60\kms\times\exp{(-R/12\kpc)}$ (i.e. we consider the effects of axisymmetric drift). The vertical velocity has a mean value of $\overline{v}_z=0 \kms$, and a standard deviation of $\sigma_{v_z}=45\kms\times\exp{(-R/20\kpc)}$.} We will refer to this particular set of initial conditions as the standard version hereafter. 

{We find that the number of test particles included in the $\lv$ diagram has little impact on the resulting envelope when the sample size is sufficiently larger ($\sim 10,000$). To ensure a fair comparison with the observational data, we still employ the same number of test particles as the observational maser stars to generate the $\lv$ envelope.}

\begin{figure*}[ht]
  \centering
  \includegraphics[width=0.9\textwidth]{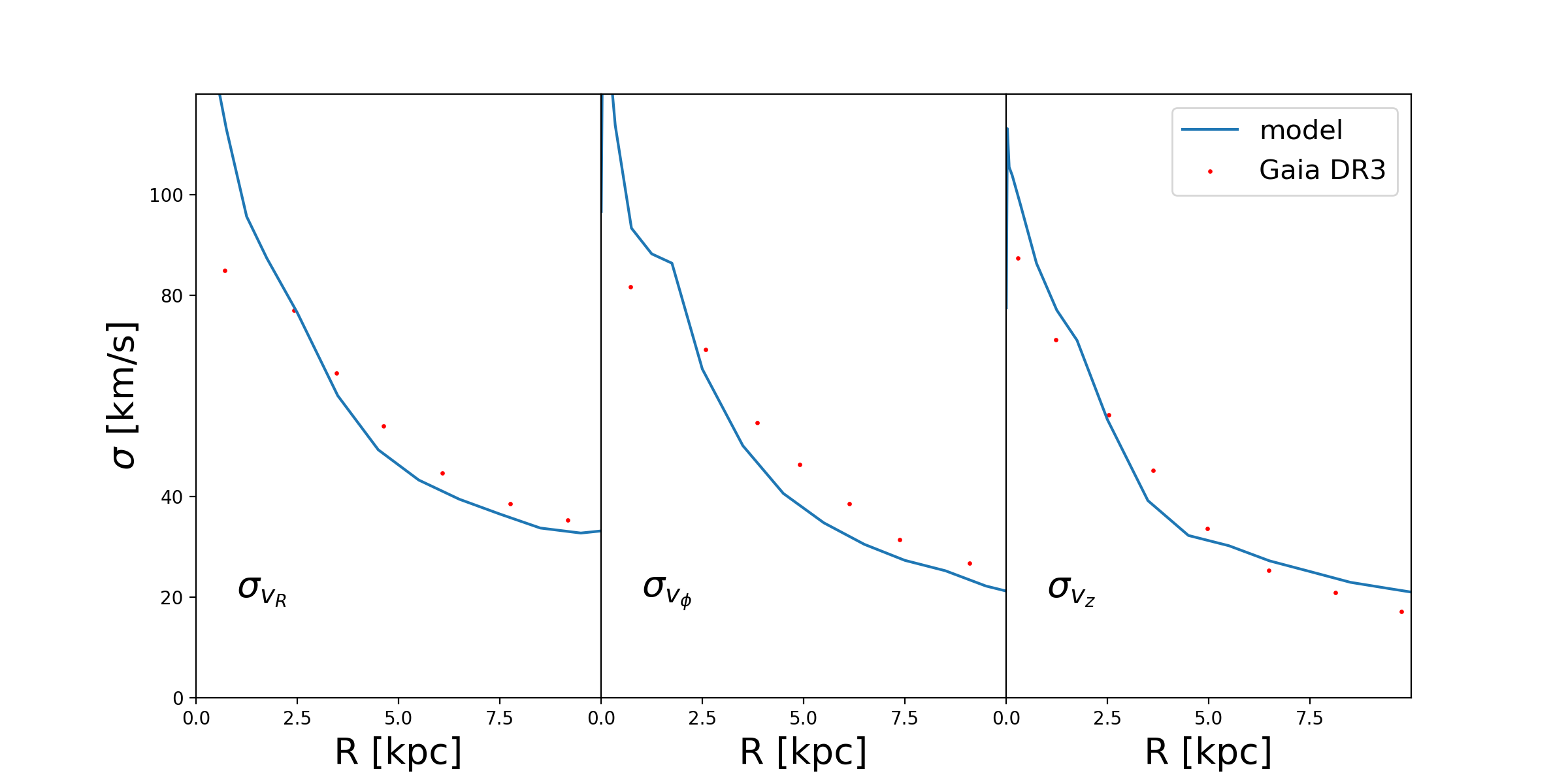}\\

  \caption{The velocity dispersion profiles in $R$, $\phi$ and $z$ directions after integration. The red points are extracted from \citet{gaiadr3}. The blue curves are the distribution obtained from integrating test particles in the \citetalias{P17} potential.}
  \label{fig:sigma_v}
\end{figure*}

To verify the accuracy of the velocity dispersion of the test particles in reflecting the observational data, we present the velocity dispersion profiles after integration in Figure~\ref{fig:sigma_v}. The integration time is set to be $10\Gyr$ and is considered sufficiently long as we demonstrate that the kinematics of test particles after integration have converged. The kinematics of our test particles well reproduce the velocity dispersion distribution observed in the Gaia DR3 data \citep{gaiadr3} in the outer regions, while we notice a mismatch at $R<2.5\kpc$. Nevertheless, our analysis in \S~\ref{section:sigma} reveals that the setting of the initial velocity dispersions in the outer region has little effect on the central envelope. In addition, the Gaia data measurement may lose accuracy at the Galactic center. Therefore, we think these discrepancies are acceptable.

\subsection{Maser observation compared with models}

Finally, we use the test particles integrated from the initial conditions of the standard version to compare with the observational data.

To estimate the error bar of the envelope, we employ the following bootstrap method: \romannumeral1) We generate 150,000 test particles using the parameters described in \S~\ref{section: initialcondition} and integrate them for a sufficiently long time to obtain their final phase space locations. \romannumeral2) We randomly select 15,207 test particles from the total sample after integration. \romannumeral3) We calculate the longitude and LOS velocities of the test particles and follow the process described in \S~\ref{section:lv_define} to define the envelope. \romannumeral4) We repeat the procedures \romannumeral2) and \romannumeral3) for 100 times. \romannumeral5) We define the median value of envelope velocities in each longitude bin as the envelope value and the range between $25\%$ and $75\%$ quantiles as the error bar. 

In addition to the bootstrap method, we also test a Monte-Carlo algorithm of obtaining the $\lv$ envelope. The main difference lies in the generation of test particles. In the Monte-Carlo algorithm, we follow the process in \S~\ref{section: initialcondition} to generate 15,207 test particles for 100 iterations. We then integrate each test particle and obtain the median envelope as well as the error bar, similar to the bootstrap method. The final results obtained using the Monte Carlo algorithm are very similar to those obtained with the bootstrap method.  

{Note that we make sure that the longitude distribution and the number of test particles are consistent with the maser observation in the procedure \romannumeral2) of \S~\ref{section: initialcondition}. We also test the results of randomly selecting particles without considering the longitude distribution and/or the number of the observed masers, and find little difference in the final envelopes resulted from the two procedures. }

\begin{figure}[htb]
  \centering
  \includegraphics[width=0.44\textwidth]{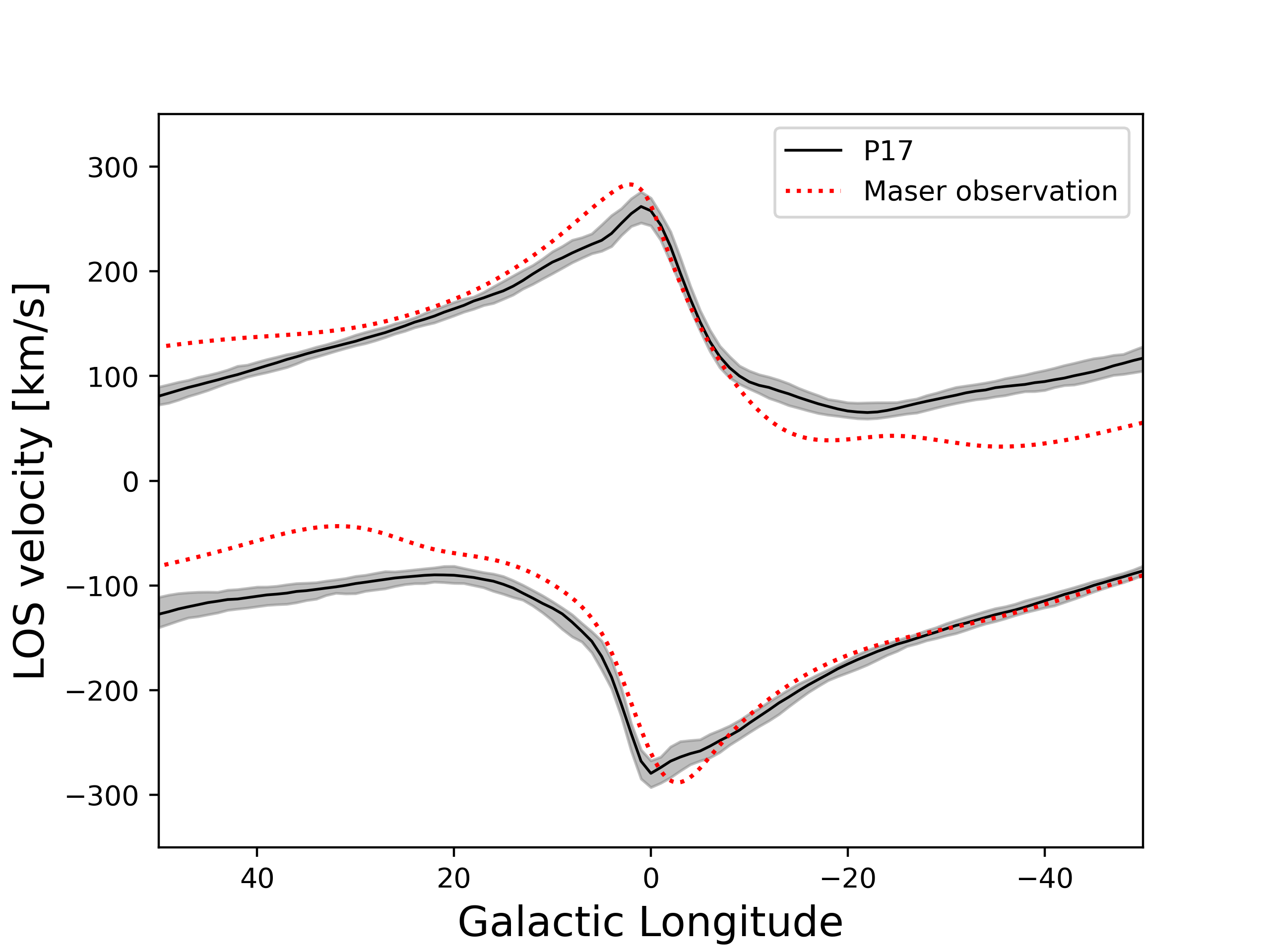}\\
  \caption{The comparison of the $\lv$ envelope between the observations and the \citetalias{P17} model. The solid black line represents the $\lv$ envelope derived from the \citetalias{P17} model, and the shaded region denotes the error bar. The maser envelope is shown in the red dashed line.}
  \label{fig:compare_data}
\end{figure}

Figure~\ref{fig:compare_data} shows the final $\lv$ envelope derived from the \citetalias{P17} potential. In general, the model and data are roughly consistent with each other, and the velocity dispersion profiles of test particles also nicely match Gaia DR3. This indicates that the \citetalias{P17} potential can largely trace the kinematics of SiO masers. However, small discrepancies are still obvious in Figure~\ref{fig:compare_data}. Specifically, the envelope in the model appears to be slightly more extended in Quadrants A and C while being slightly less extended in Quadrant B.



Moreover, a central asymmetry difference is observed between the model and data in Quadrants B and D of the envelope. To be more specific, the model generates a smaller envelope in Quadrant B,  but a comparable one in Quadrant D compared with the observational data. The detailed discussion of these discrepancies can be found in Appendix~\ref{section: asymmetry}.

\section{Factors affecting the $\lv$ envelope}
\label{section:factor}

In this section, we investigate the factors that influence the $\lv$ envelope and aim to gain insights into how the $\lv$ envelope can constrain the bar potential.

We begin with the analysis of the distribution of different orbits in the $\lv$ diagram.

Considering a star on a circular orbit with a radius $R$ and a longitude $l$, the LOS velocity  $v_{\rm{LOS}}$ of the star can be expressed as:
\begin{equation}
 v_{\rm{LOS}}=[\Omega(R)-\Omega_0]R_0\sin{l},
\label{eq:circular_vl}
\end{equation}
where $R_0$ is the solar radius, $\Omega_0$ is the angular frequency of the circular motion at the solar radius, and $\Omega(R)$ represents the angular frequency of the circular motion at the cylindrical radius $R$ \citep{1998Binney}.

According to Equation~\ref{eq:circular_vl}, stars on circular orbits outside the solar radius are located in Quadrants A and C of the $\lv$ diagram. Conversely, stars on circular orbits inside the solar radius are found only in Quadrants B and D. Quadrants A and C of the $\lv$ diagram are commonly referred to as the 'forbidden' regions in gas dynamics, as gas in these regions are not expected if assuming circular motions within the solar radius. 

In Appendix~\ref{appendix: disk}, we test circular orbits in an axisymmetric disk potential. The results demonstrate that all circular orbits are distributed in Quadrants B and D. This finding is consistent with the conclusion derived from Equation~\ref{eq:circular_vl}.

\begin{figure*}[htbp!]
  \centering
  \includegraphics[width=0.9\textwidth]{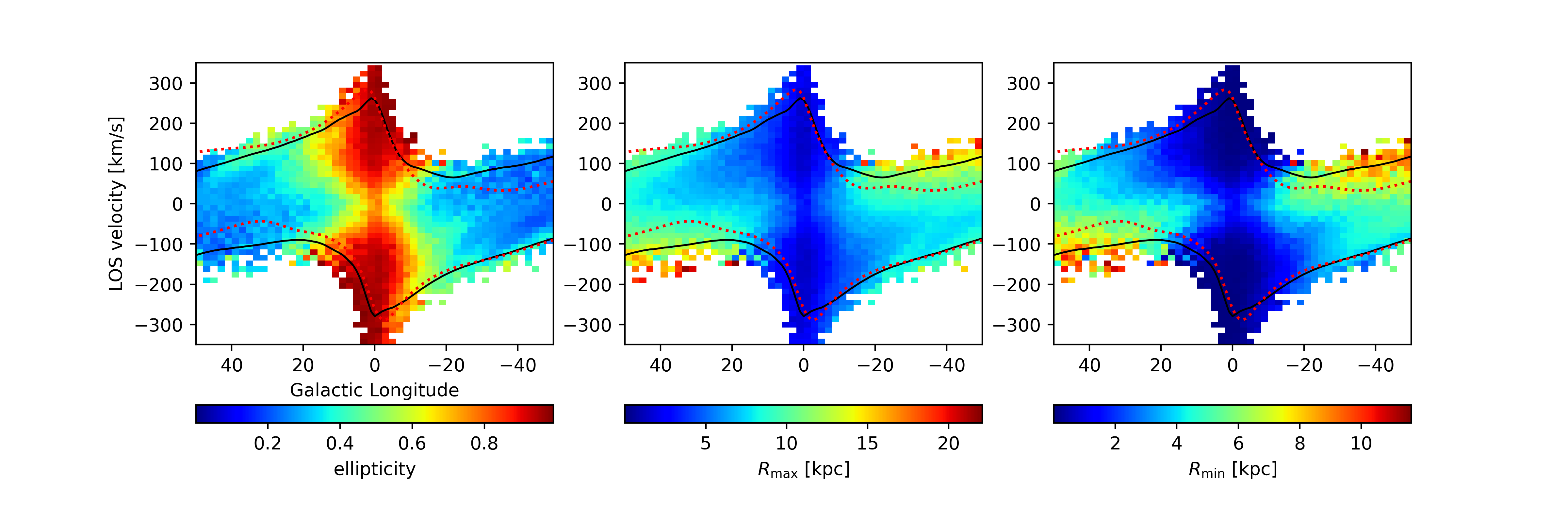}\\

  \caption{The orbital properties of test particles in the $\lv$ diagram. In all three panels, the red dashed lines represent the envelopes of the observational data, while the black lines represent the envelopes generated by the model. The left panel is color-coded by ellipticity, defined by $(R_{\rm{max}}-R_{\rm{min}})/(R_{\rm{max}}+R_{\rm{min}})$, where $R_{\rm{max}}$ and $R_{\rm{min}}$ are the maximum and minimum radius of the orbits which the given test particle belongs to. The middle and right panels are color-coded by $R_{\rm{max}}$ and $R_{\rm{min}}$, respectively.}
  \label{fig:cir_lv}
\end{figure*}

Next, we extend our analysis to include non-circular orbits in the bar potential. {As shown in Figure~\ref{fig:cir_lv}, stars on non-circular orbits (yellow and red bins in the left panel of Figure~\ref{fig:cir_lv}) tend to lie in the central region ($|l| \lesssim 20^\circ$), which is the region influenced primarily by the Galactic bar. In contrast, stars on nearly-circular orbits (blue bins in the left panel) predominantly occupy the outer region ($|l| \gtrsim 30^\circ$) of the $\lv$ diagram. We have further conducted tests on the distribution of the maximum and minimum radii of the orbits in the $\lv$ diagram in the middle and right panels of Figure~\ref{fig:cir_lv}. Clearly, orbits outside the solar radius (yellow and red bins in the middle and right panel of Figure~\ref{fig:cir_lv}) are distributed in Quadrants A and C, and nearly-circular orbits within the solar radius are located in Quadrants B and D. This also aligns with the implications of Equation~\ref{eq:circular_vl}.}


Once the distribution of orbital properties in the $\lv$ diagram has been examined, it is possible to investigate the various mechanisms that influence the shape and characteristics of the $\lv$ envelope. These factors can be broadly classified into two categories: those associated with observational factors and those associated with model settings.

\subsection{Observational factors}
\label{section: selection function}

The primary observational factor that significantly impacts the $\lv$ envelope is the heliocentric selection function. The maser data do not sample the far side as well as the near side in the maser data. This indicates that observations are not complete enough to capture all stars in the bulge regions, especially those on the far side \citep{maser_distance_2024}. As a result, the resulting $\lv$ envelope may not accurately represent the full extent of the stellar distribution in the bulge, in particular those in Quadrants A and D.

To mimic the effect of the heliocentric selection function, we apply a direct heliocentric cut, effectively creating a volume-limited sample. Additionally, we examine the effects of a radially-dependent selection function, which is a Gaussian-like function centered on the Sun and reduces the probability of detecting stars on the far side. These tests yield similar conclusions. 

\begin{figure*}[htbp!]
  \centering
  \includegraphics[width=0.44\textwidth]{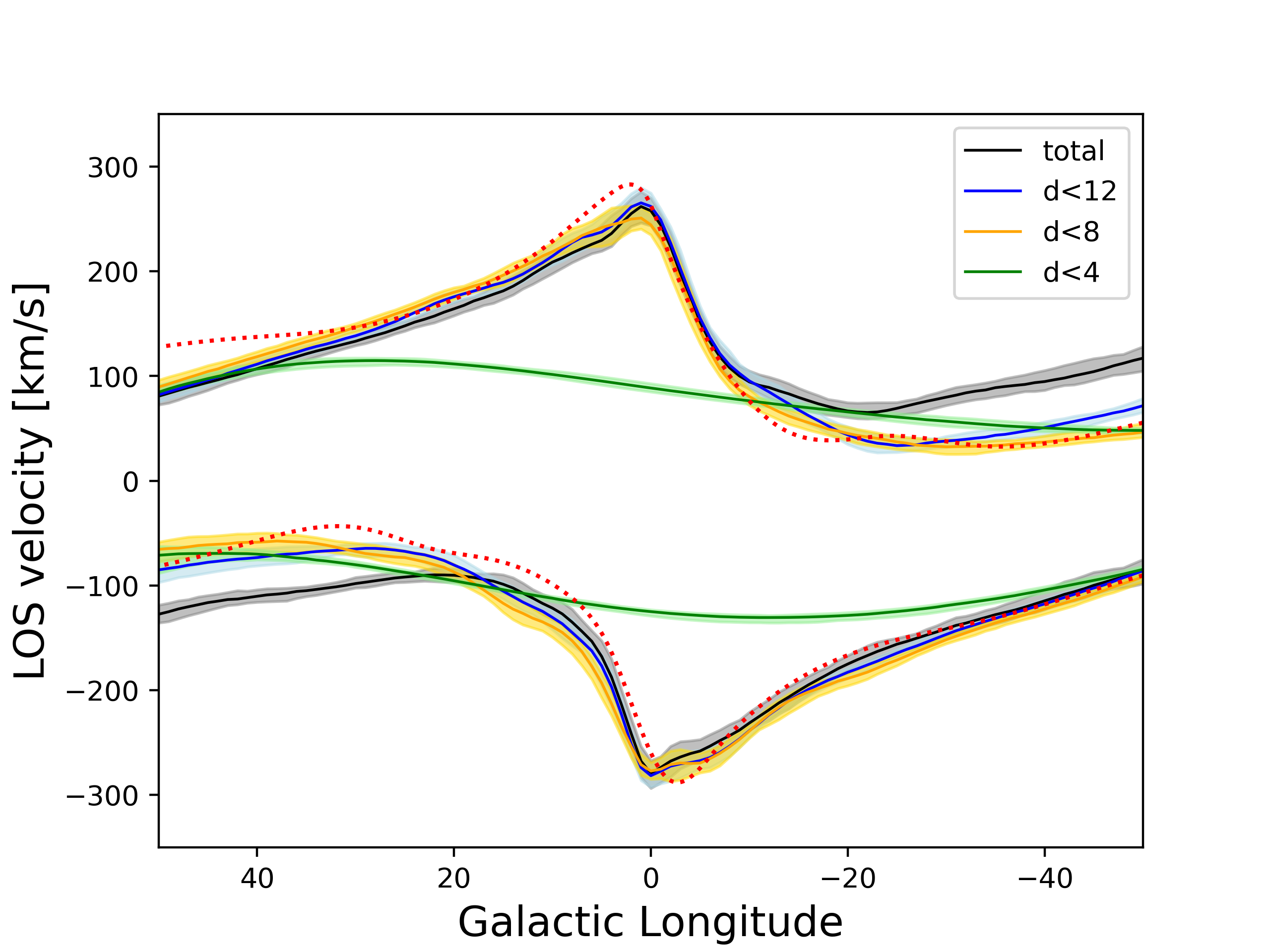}
  \includegraphics[width=0.44\textwidth]{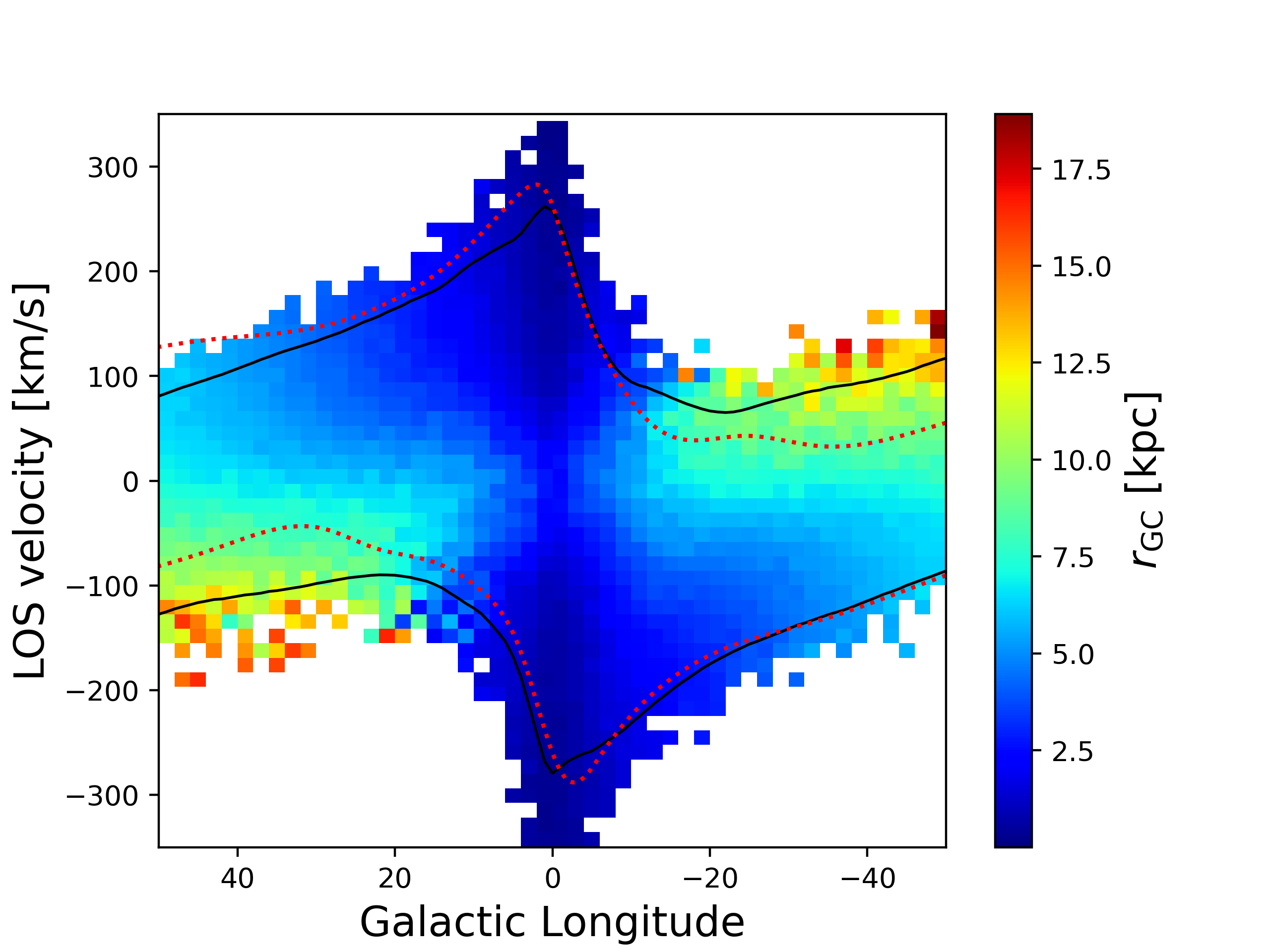}
  \caption{ The effect of heliocentric selection functions on the $\lv$ envelope. The left panel shows the different $\lv$ envelopes of applying different heliocentric cuts. {The black curve labeled 'total' corresponds to the standard version of the model, with no heliocentric selection function applied.} 
  The shaded regions represent the error bar. The right panel shows the $\lv$ diagram color-coded by the distance between the test particles and the Galactic center, which offers an explanation for the variation of envelopes of different heliocentric cuts. The red dashed lines in both panels represent the maser envelope.}
  \label{fig:helio}
\end{figure*}

As shown in the left panel of Figure~\ref{fig:helio}, applying a relatively large heliocentric cut ($d_{\rm{cut}}>8 \kpc$) leads to a decrease in the outer regions of the envelope in Quadrants A and C. Conversely, applying a relatively small heliocentric cut ($d_{\rm{cut}}<8 \kpc$) results in a drop in the central region of the envelope, possibly due to excluding most of the stars within the bar region.

The impact of the heliocentric selection function on Quadrants A and C of the $\lv$ envelope can be elaborated by the right panel of Figure~\ref{fig:helio}. Orbits with relatively large radii ($R>8 \kpc$) predominantly occupy Quadrants A and C of the $\lv$ diagram. When a relatively large heliocentric selection cut is applied, there is a significant reduction in the probability of detecting these orbits at large radii, particularly those outside the solar radius. However, other orbits with relatively small radii may still be detected. Consequently, the drop in the envelope is primarily observed in Quadrants A and C.


The data we use is limited to latitude $|b|\lesssim 6\degree$. We have verified that latitude restriction has virtually no impact on the $\lv$ envelope.


\subsection{Model setting factors}

In this section, we investigate the impact of various model settings on the properties of the $\lv$ envelope. Three key factors are considered to be the most influential: the initial kinematic properties of the test particles, the monopole term of the bar potential, and the quadrupole term of the bar potential. Additionally, we briefly examine a few other potential factors that may have a relatively minor impact on the properties of the envelope.

\subsubsection{Initial kinematic properties of the test particles}
\label{section:sigma}

\begin{figure}[htbp!]
  \centering
  \includegraphics[width=0.44\textwidth]{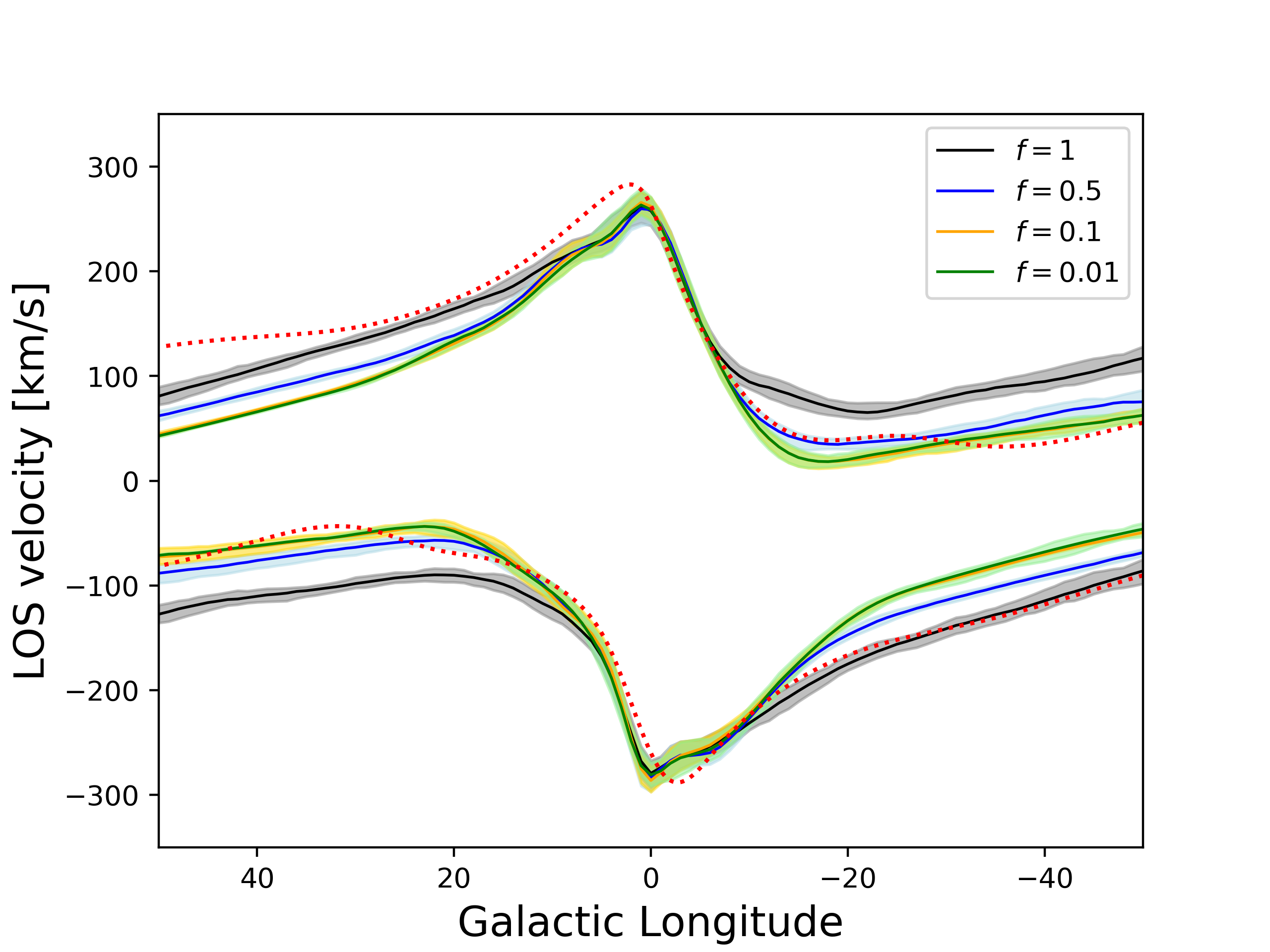}\\
  \caption{The different $\lv$ envelopes of models with different initial velocity dispersions.  $f=1$, $0.5$, $0.1$, and $0.01$ represent that models have the initial velocity dispersions multiplying the standard initial velocity dispersions by $1$, $0.5$, $0.1$, and $0.01$, respectively. The shaded regions represent the error bar. The red dashed line represents the maser envelope.}
  \label{fig:ini_v_tot}
\end{figure}

In Figure~\ref{fig:ini_v_tot}, we compare different $\lv$ envelopes of models with different initial velocity dispersions. We define $f$ as the relative factor of the velocity dispersions of the standard initial conditions compared to the velocity dispersions of each model. $f=1$, $0.5$, $0.1$, and $0.01$ indicate that models have the initial velocity dispersions multiplying the standard version by $1$, $0.5$, $0.1$, and $0.01$, respectively. We consider $f=1$ as the upper limit since the model of the standard version has already exhibited more extended envelopes in Quadrants A and C.



{We find that the initial velocity dispersion has a significant impact on the envelope. This is because varying velocity dispersions can lead to varying proportions of circular and non-circular orbits, influencing the overall shape of the $\lv$ envelope. However, the initial velocity dispersion may no longer affect the envelope if it is sufficiently small ($f<0.1$), as the velocity dispersion of particles after orbital integration converges in that case.}

{Note that the variation of the envelopes mainly lies in the outer region of the $\lv$ diagram. This is because particles with different initial kinematic properties tend to achieve a similar velocity dispersion profile within a small radius ($\sim 2 \kpc$) after orbital integration. The changes of the $\lv$ envelopes in Figure~\ref{fig:ini_v_tot} are indeed driven by the velocity dispersion profiles after time-integrating the particles.}

\begin{figure*}[htbp!]
  \centering
  \includegraphics[width=0.3\textwidth]{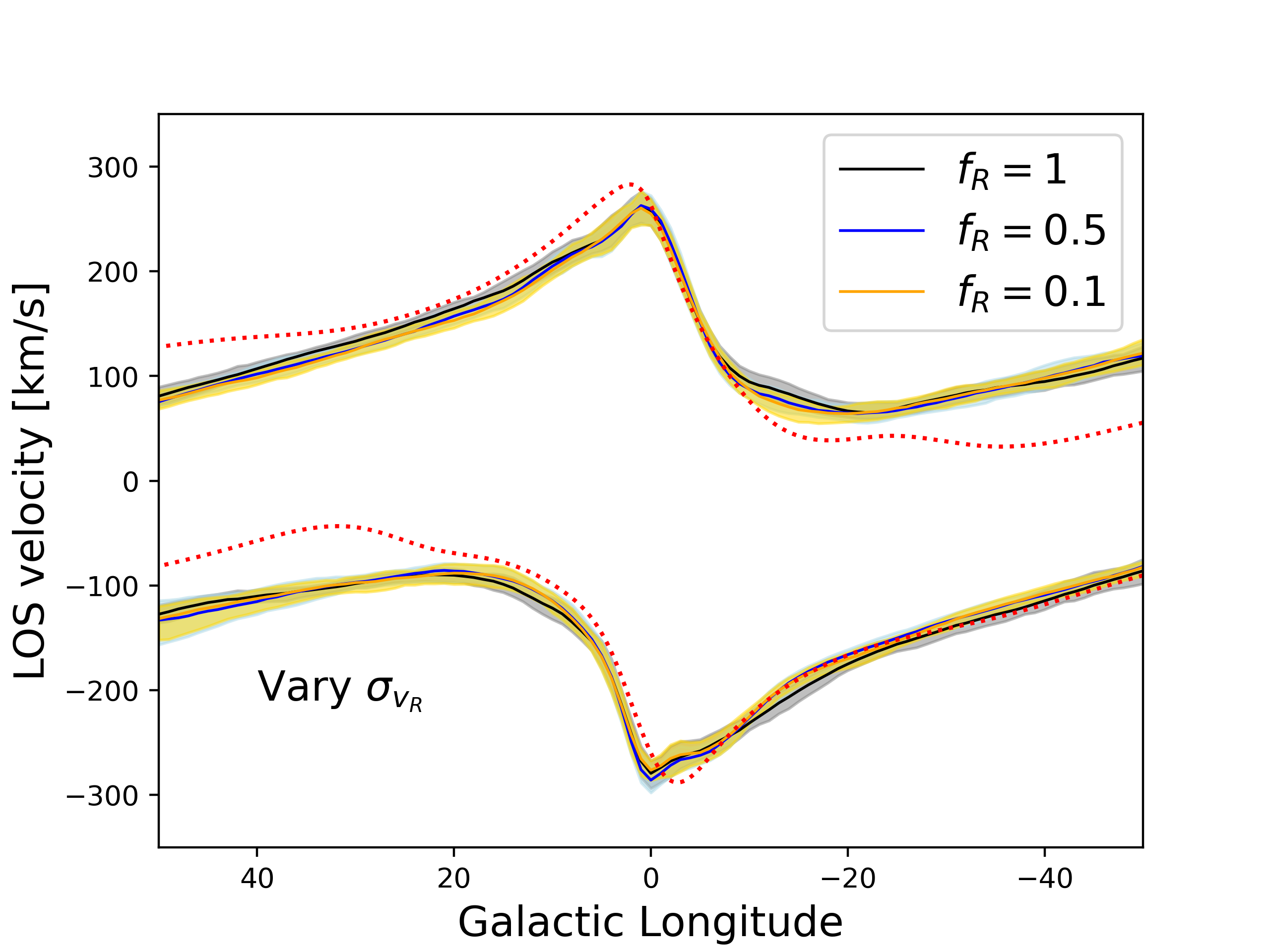}
  \includegraphics[width=0.3\textwidth]{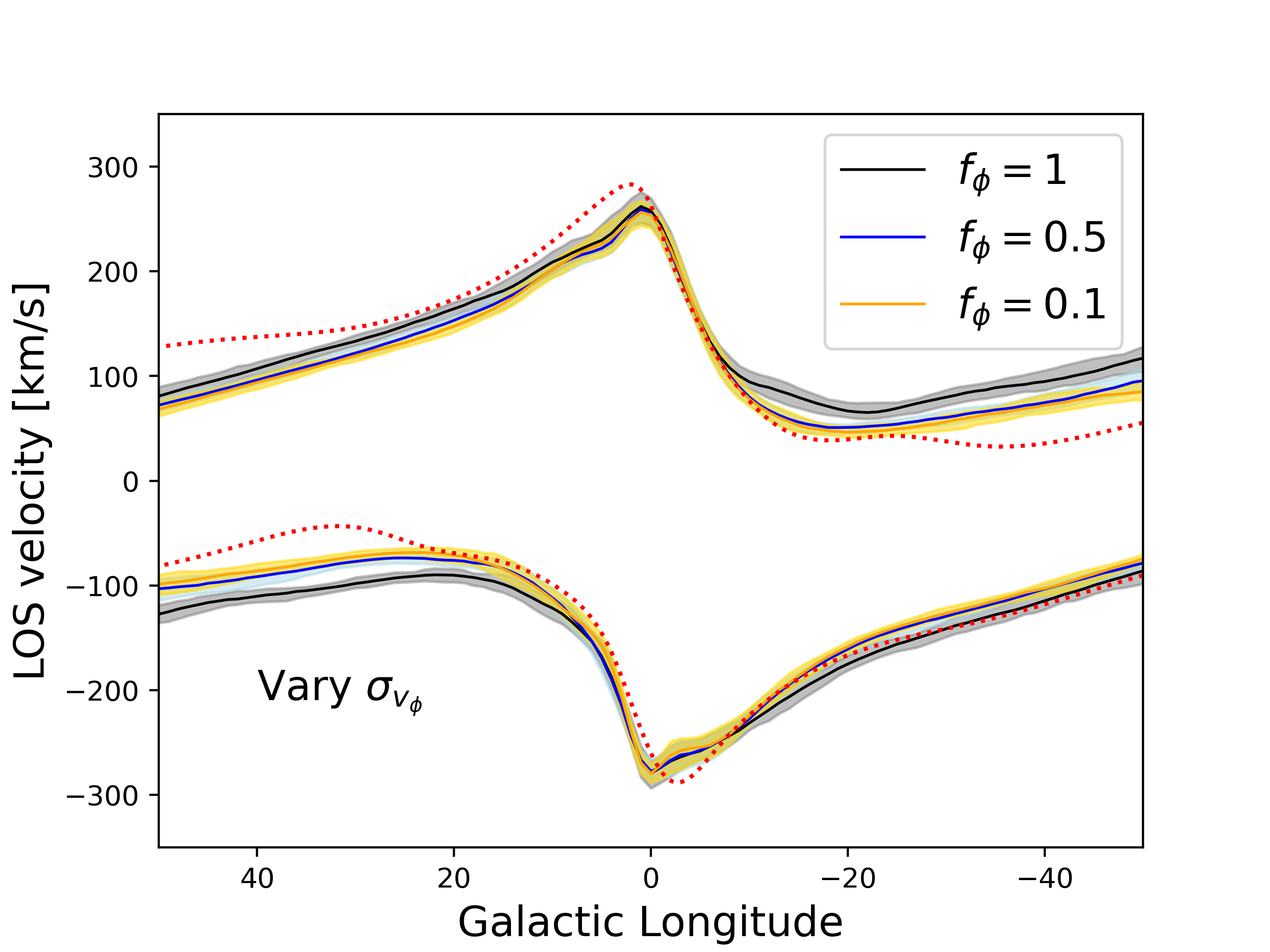}
  \includegraphics[width=0.3\textwidth]{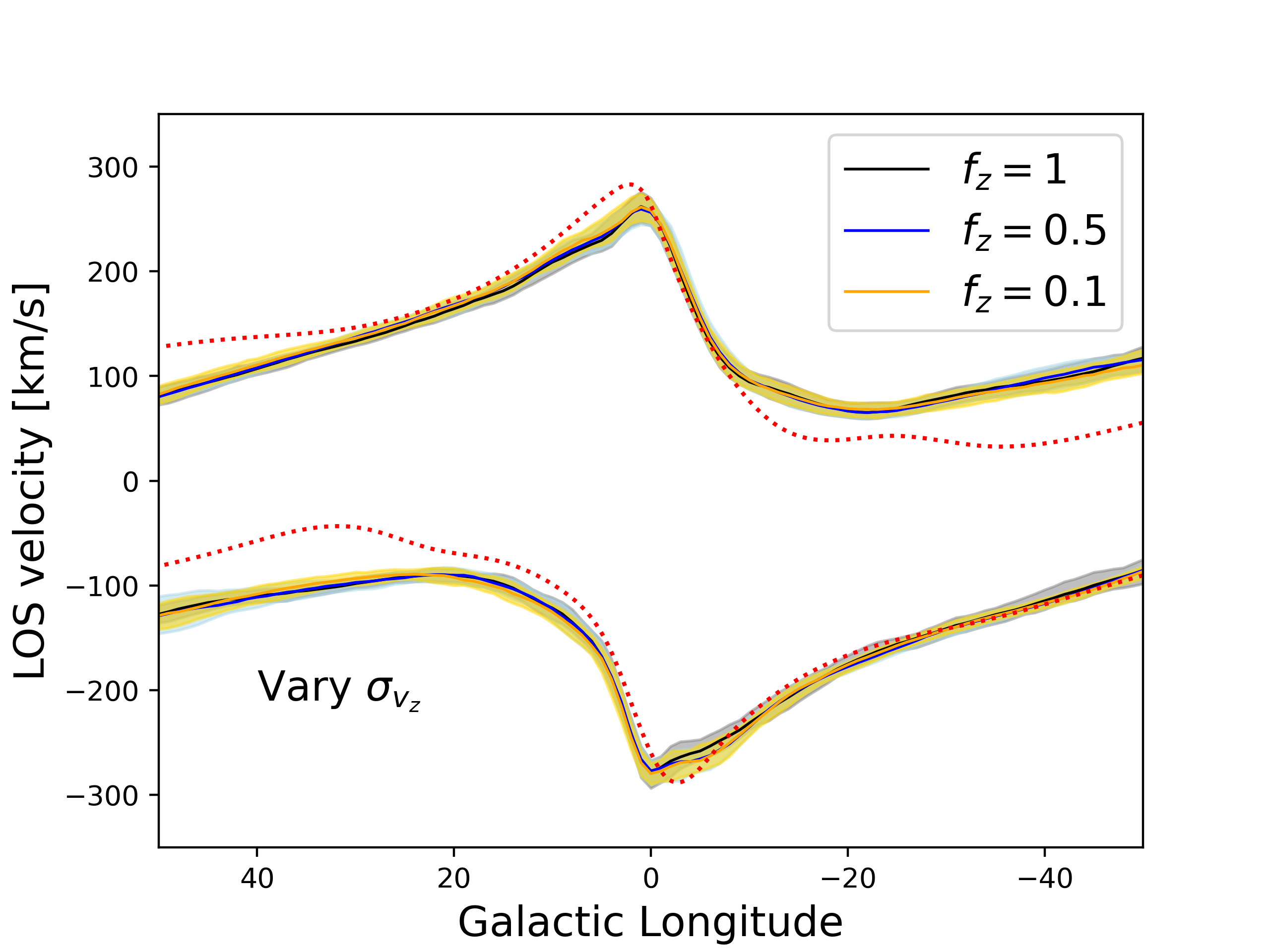}
  \caption{The different $\lv$ envelopes of varying the initial velocity dispersions in different directions. To be specific, the left, middle and right panels show the resulted envelopes if varying the initial velocity dispersions in $R$, $\phi$ and $z$ direction, respectively. The shaded regions represent the error bar. The red dashed lines in all three panels represent the maser envelope. }
  \label{fig:ini_v_dirction}
\end{figure*}

We discuss the similar phenomenon observed in the case of the axisymmetric disk potential in Appendix~\ref{appendix: disk}. When the initial velocity dispersion decreases, the proportion of non-circular orbits diminishes, i.e., no signal in Quadrants A and C as expected. Consequently, the resulting $\lv$ envelope moves closer to the envelope generated solely by circular orbits, which are predominantly distributed in Quadrants B and D of the $\lv$ diagram. 

Furthermore, our analysis reveals that the influence of varying the initial velocity dispersion can be decomposed into different directions. We define $f_{i}$, where $i=R,\phi,z$, as the relative factor of the standard initial velocity dispersions in $i$ direction compared to the initial velocity dispersions the model holds in $i$ direction. $f_i=1$, $0.5$, $0.1$ indicate that the models have the initial velocity dispersions in the $i$ direction multiplying the standard version by $1$, $0.5$, and $0.1$, respectively, while the components in other directions remain the same as the standard version.

Figure~\ref{fig:ini_v_dirction} demonstrates that alterations in the envelope resulting from changes in the initial velocity dispersion are primarily driven by variations in $\sigma_{v_\phi}$ (middle panel). This is attributed to the fact that non-circular motions in the bar potential is primarily affected by the velocity dispersions in the tangential and radial direction. In addition, the variation of $\sigma_{v_R}$ also changes $\overline{v}_\phi$ due to the effect of asymmetric drift. Since $\overline{v}_\phi$ influence the $\lv$ envelope in an opposite manner compared with $\sigma_{v_R}$, their combined effect on the envelope is negligible. Consequently, initial velocity dispersion in the tangential direction becomes the main factor that influences the $\lv$ envelope.


In addition, we have conducted an investigation into the potential impact of the initial vertical thickness on the $\lv$ envelope. Our analysis reveals that, similar to the initial vertical velocity dispersion, the initial vertical thickness has minimal influence the $\lv$ envelope. Therefore, we can conclude that the initial vertical motion is not a critical factor in determining the $\lv$ envelope.

{The variations of envelopes generated by different initial conditions suggest that the kinematics of test particles have a substantial impact on the envelope. Although we have made efforts to carefully set the initial conditions to match the kinematics of Gaia DR3 data, there remains a slight disparity for the envelopes in Quadrants A and C between the model and the maser observation. This difference may be attributed to the heliocentric selection function of the model, as discussed in \S~\ref{section: selection function}.}


\subsubsection{The monopole and quadrupole terms of the bar potential}

The multipole expansion is a commonly used method for solving the gravitational potential in Galactic dynamics. For a general bar potential, the expansion is usually dominated by the monopole and the quadrupole terms. The monopole term, representing the value of the rotation curve, characterizes the circular motion of stars within the potential. On the other hand, the quadrupole term captures important bar parameters such as the axis ratio and mass of the bar.

Our exploration reveals that the monopole and quadrupole terms have distinct impacts on the properties of the $\lv$ envelope. 

To better separate the two terms, we adopt the potential similar to \citet{Sormani2015}. The potential consists of only monopole and quadrupole terms:
\begin{equation}
 \Phi=\Phi_0(r)+\Phi_2(r,\theta)\cos{(2\phi)},
 \label{eq:sormani2015}
\end{equation}
where $(r,\theta,\phi)$ is the spherical polar coordinate. $\Phi_0(r)$ represents the monopole term and $\Phi_2(r,\theta,\phi)$ represents the quadrupole term.

Here we apply a simple logarithmic potential to $\Phi_0$:
\begin{equation}
 \Phi_0(r)=\frac{1}{2}v_0^2\ln{(R_e^2+r^2)},
 \label{eq:phi0}
\end{equation}
{where we set $v_0=220\kms $ and $R_e=0.35\kpc $, slightly modified from \citet{sormani2018} to better match the inner rotation curve of \citetalias{P17}.}

{The quadrupole term $\Phi_2$ is generated by the density distribution $\rho_2$ given by \citet{Sormani2015, sormani2018}:}
\begin{equation}
 \rho_2(r,\theta,\phi)=\frac{KA}{r_{\rm{q}}^2}\exp{(-\frac{2r}{r_{\rm{q}}})}\sin^2\theta\cos(2\phi),
 \label{eq:rho2}
\end{equation}
{where $K=\frac{v_0^2 e^2}{4\pi G}$. Quadrupole strength $A$ and quadrupole length $r_{\rm{q}}$ are the two main free parameters of $\rho_2$. In the fiducial model, we set $A=0.4$ and $r_{\rm{q}}=1.5\kpc$ as they are the best-fit parameters suggested by \citet{Sormani2015}.}

\begin{figure*}[htbp!]
  \centering
  \includegraphics[width=0.44\textwidth]{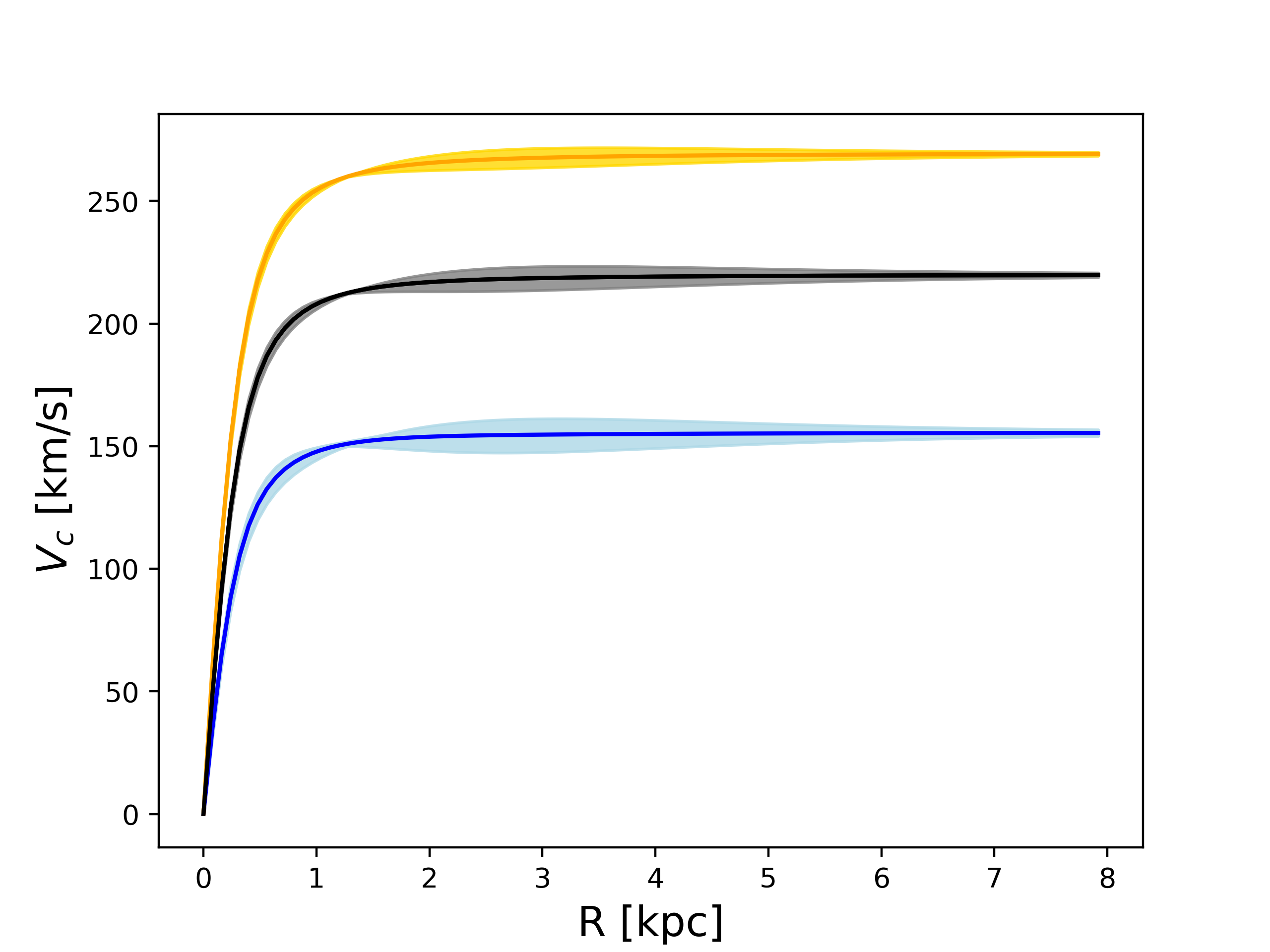}
  \includegraphics[width=0.44\textwidth]{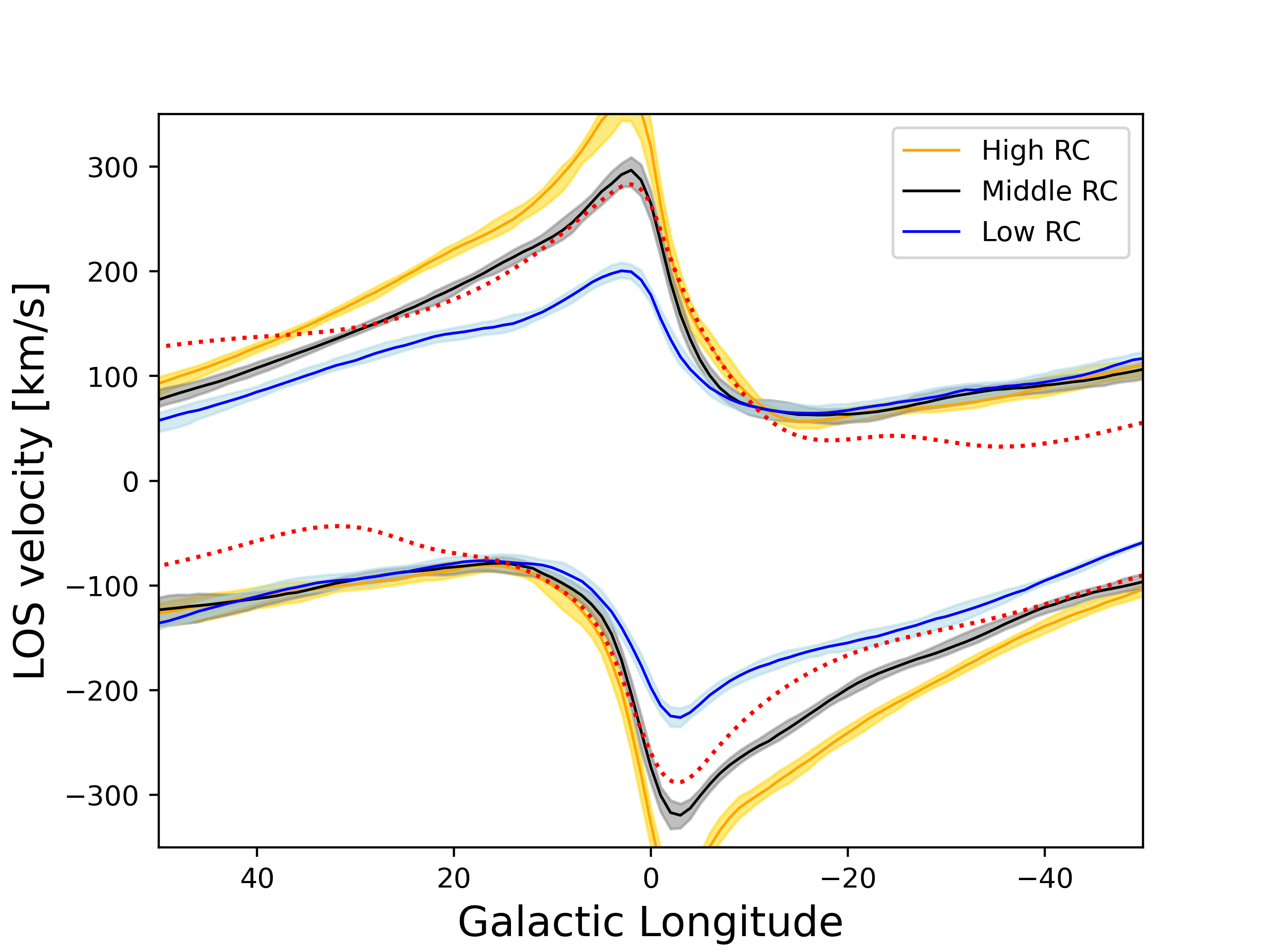}
  
  \caption{The effect of the monopole term on the rotation curves and the $\lv$ envelopes. The left panel shows the rotation curves for three models with different monopole terms. The solid line is defined by the averaged rotation velocity at the given radius, and the shaded regions are bounded by the maximum and minimum rotation velocity at the given radius. The right panel shows the envelopes generated by these three models. The shaded regions represent the error bar, and the red dashed line represents the maser envelope.}
  \label{fig:rc_enve}
\end{figure*}

We first investigated the impact of the monopole term of the potential on the $\lv$ envelope. As shown in Fig~\ref{fig:rc_enve}, the monopole term significantly affects the envelope in Quadrants B and D. Specifically, a higher monopole term, which effectively increases the value of the rotation curve,  corresponds to a larger envelope (terminal velocity) in Quadrants B and D. In Quadrants A and C, the variation of the envelopes due to changes in the monopole term only appears in the inner regions, which is aligned with that shown in Quadrants B and D. 

The impact of the monopole term on the properties of the $\lv$ envelope in Quadrants B and D can be explained by the fact that circular orbits within the solar radius are predominantly found in these regions. Therefore, changes in the rotation curve, which mainly affect the velocity of circular orbits, naturally affect the properties of the envelope there. The difference of the envelopes in Quadrants A and C, on the other hand, is likely due to changes in the distribution of non-circular orbits, such as the $x_1$ orbits, resulting from variations in the monopole term of the potential.

\begin{figure*}[htbp!]
  \centering
  \includegraphics[width=0.44\textwidth]{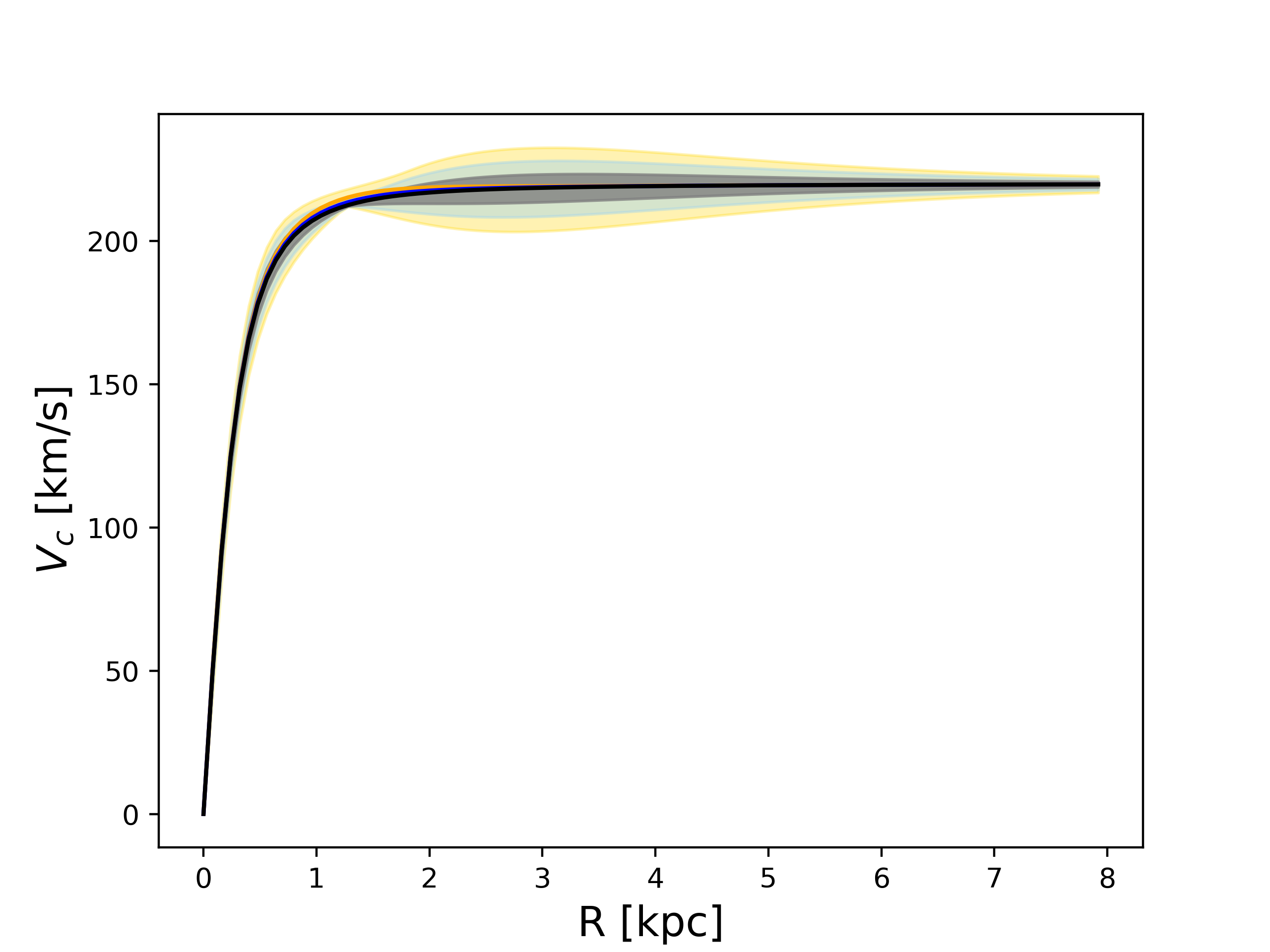}
  \includegraphics[width=0.44\textwidth]{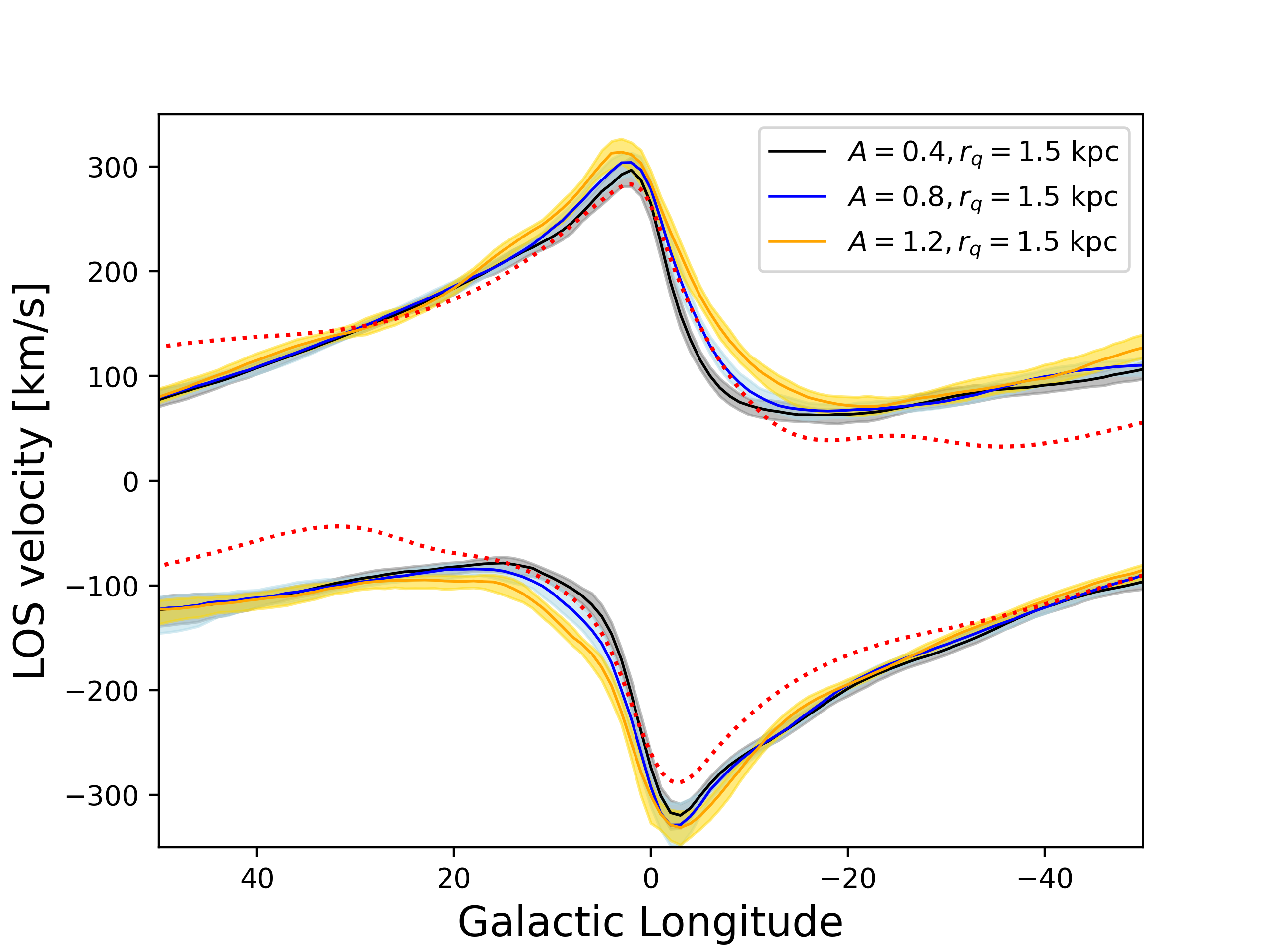}\\
  \includegraphics[width=0.44\textwidth]{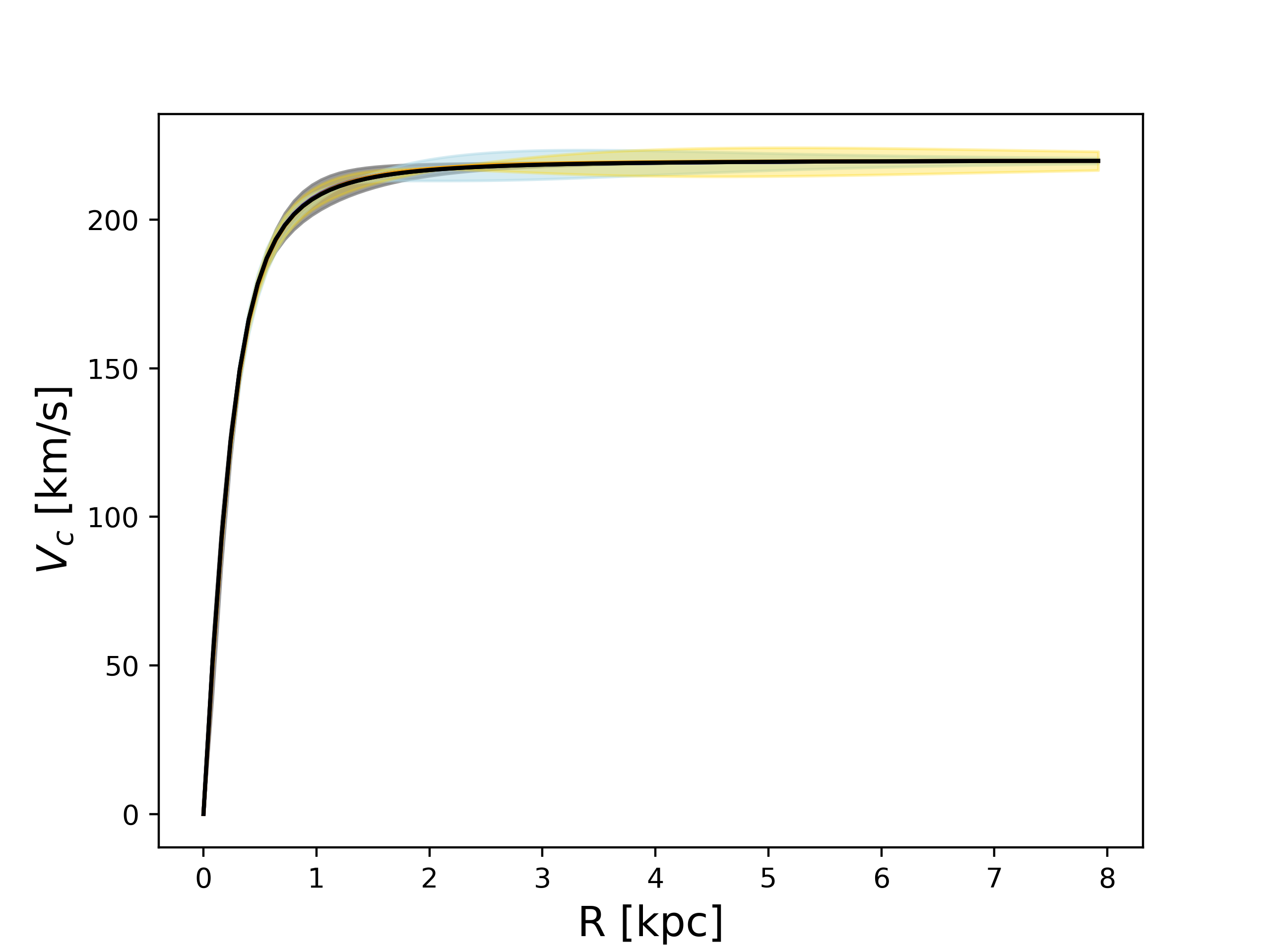}
  \includegraphics[width=0.44\textwidth]{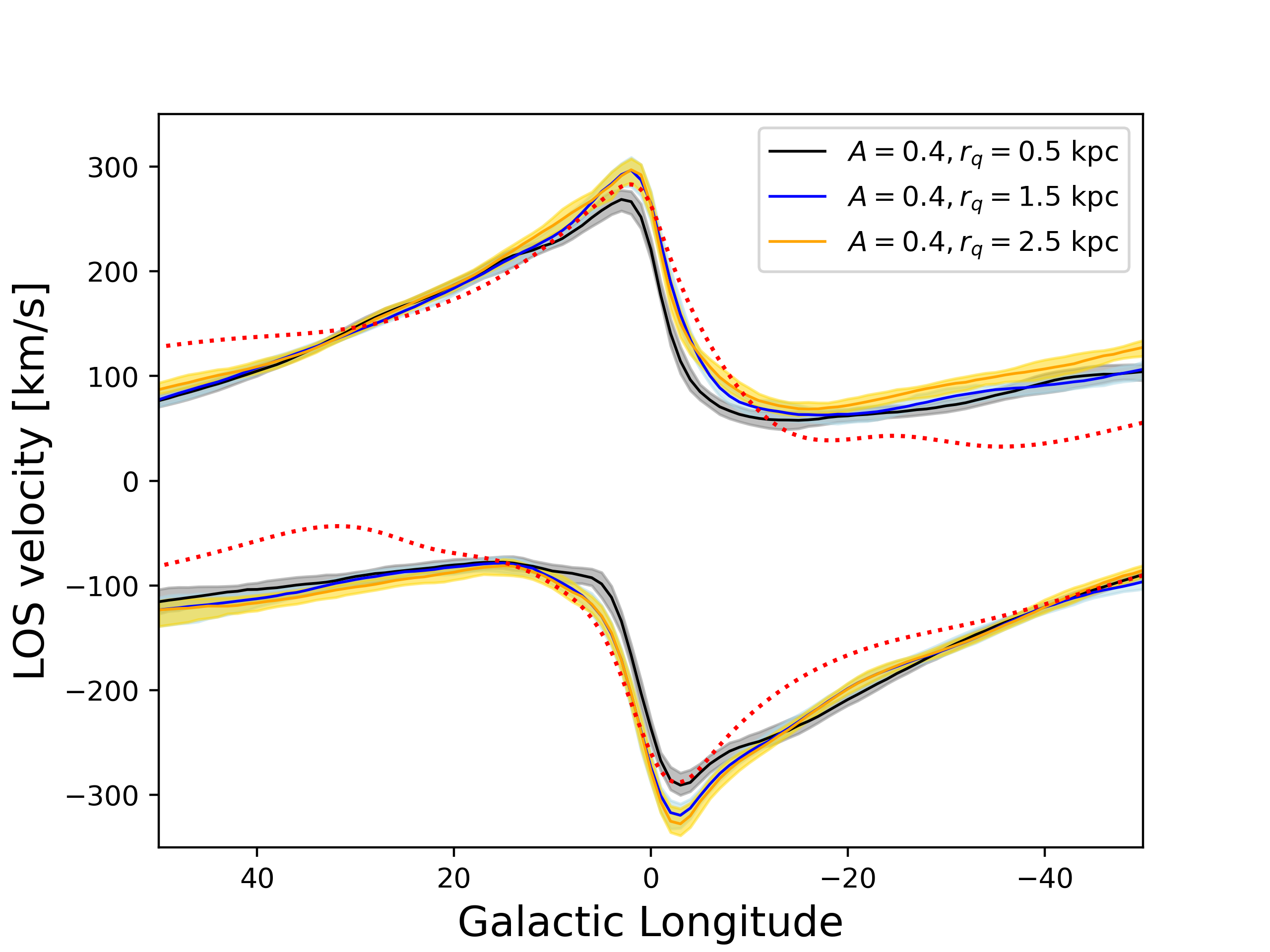}\\
  \includegraphics[width=0.44\textwidth]{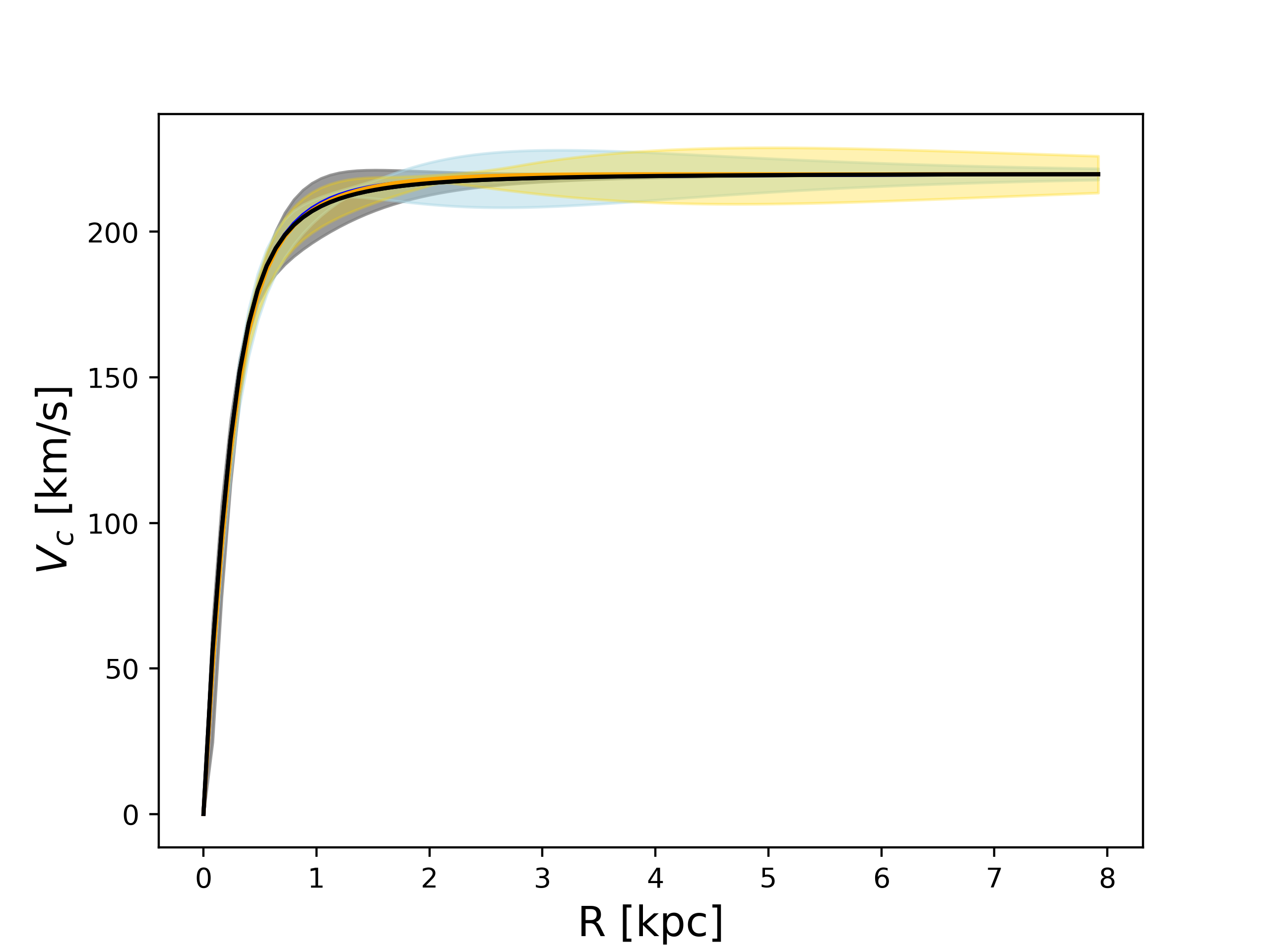}
  \includegraphics[width=0.44\textwidth]{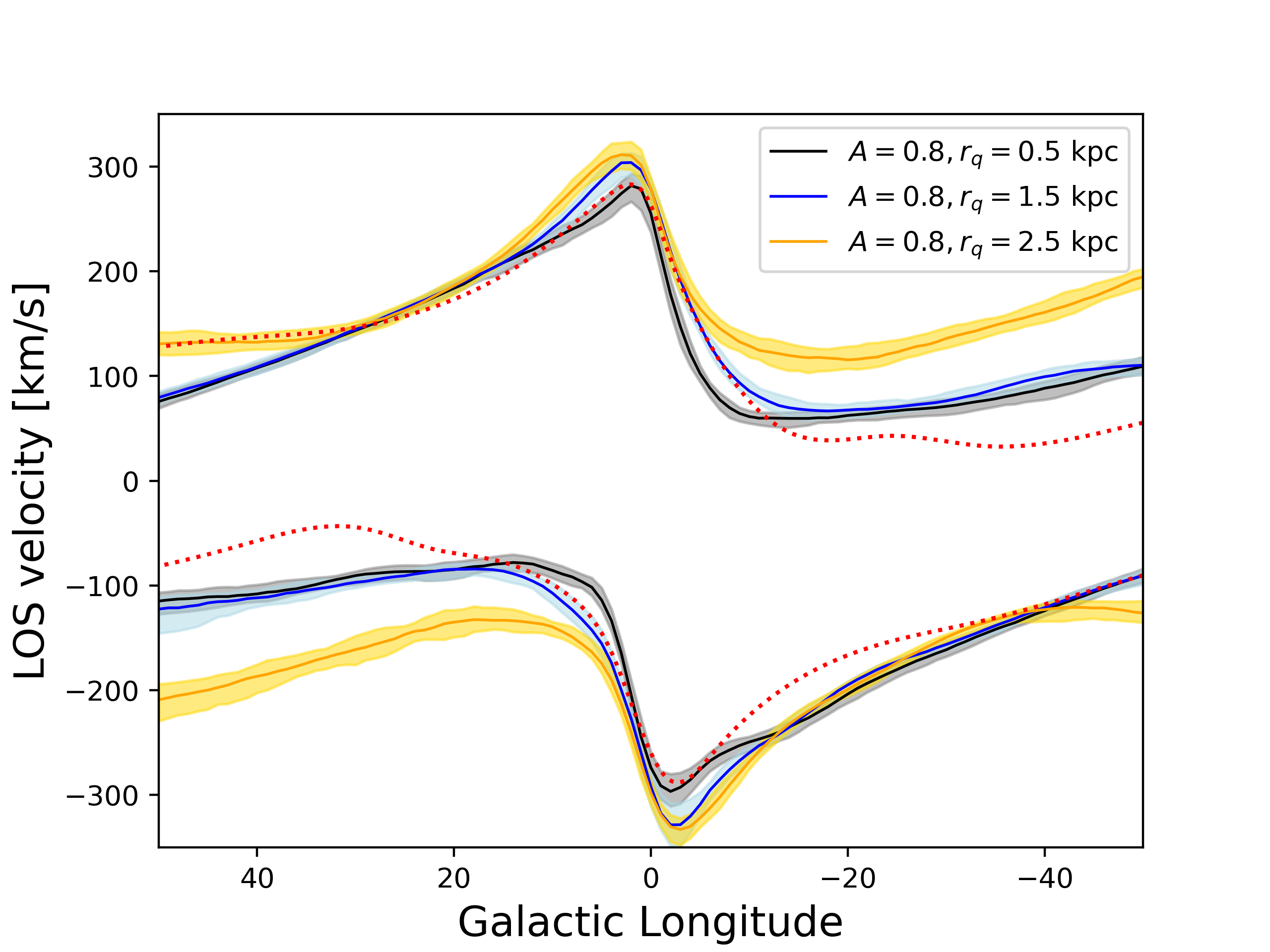}\\
  
  \caption{The effect of the varying the quadrupole term on the rotation curve and the $\lv$ envelope. In the top row, the quadrupole term is varied by only changing the quadrupole strength $A$ and keeping the quadrupole length $r_{\rm{q}}$ unchanged. The middle and lower rows only vary the quadrupole length $r_{\rm{q}}$ and fix the quadrupole strength for two values of $A$. The red dashed line in all panels represent the maser envelopes. The shaded regions in the left panels are bounded by the maximum and minimum rotation velocity at the given radius, and the shaded regions in the right panels represent the error bar.  }
  \label{fig:quad_enve}
\end{figure*}

Next, we discuss the effect caused by the quadrupole term. As shown in Figure~\ref{fig:quad_enve}, the quadrupole term mainly affects the $\lv$ envelope in Quadrants A and C, due to the fact that the quadrupole term tends to act on the distribution of non-circular orbits. {Notably, we observe slight envelope fluctuations in the central areas of Quadrants B and D, which could be due to the existence of bar-like orbits within those regions.  }

To better understand the impact of the quadrupole term on the properties of the $\lv$ envelope, we separately investigated the effects of the two free parameters in the quadrupole term: the quadrupole strength $A$ and the quadrupole length $r_{\rm{q}}$. Our analysis reveals that $A$ primarily affects the envelope in the small $|l|$ regions. By contrast, $r_{\rm{q}}$ can affect the properties of the envelope in the outer regions of the $\lv$ diagram, particularly when the fixed value of $A$ is large. Moreover, the properties of the envelope in the inner region of Quadrants A and C exhibit an approximately linear relation with $A$, but a clearly nonlinear relation with $r_{\rm{q}}$.

Our best-fit parameters in the quadrupole term of the potential are $A=0.4$ and {$r_{\rm{q}}=1.5\kpc$}. \citet{Sormani2015} suggested that $A$ should be larger than $0.4$ and $r_{\rm{q}}$ should be larger than {$1.5\kpc$} to match the Galactic bar with their hydrodynamic simulation. However, according to Figure~\ref{fig:quad_enve}, when $A$ is large ($A=0.8$), a larger value of $r_{\rm{q}}$ can also cause a significant flare at the outer region of Quadrants A and C.

\subsubsection{Other parameters}

\begin{figure}[htbp]
  \centering
  \includegraphics[width=0.44\textwidth]{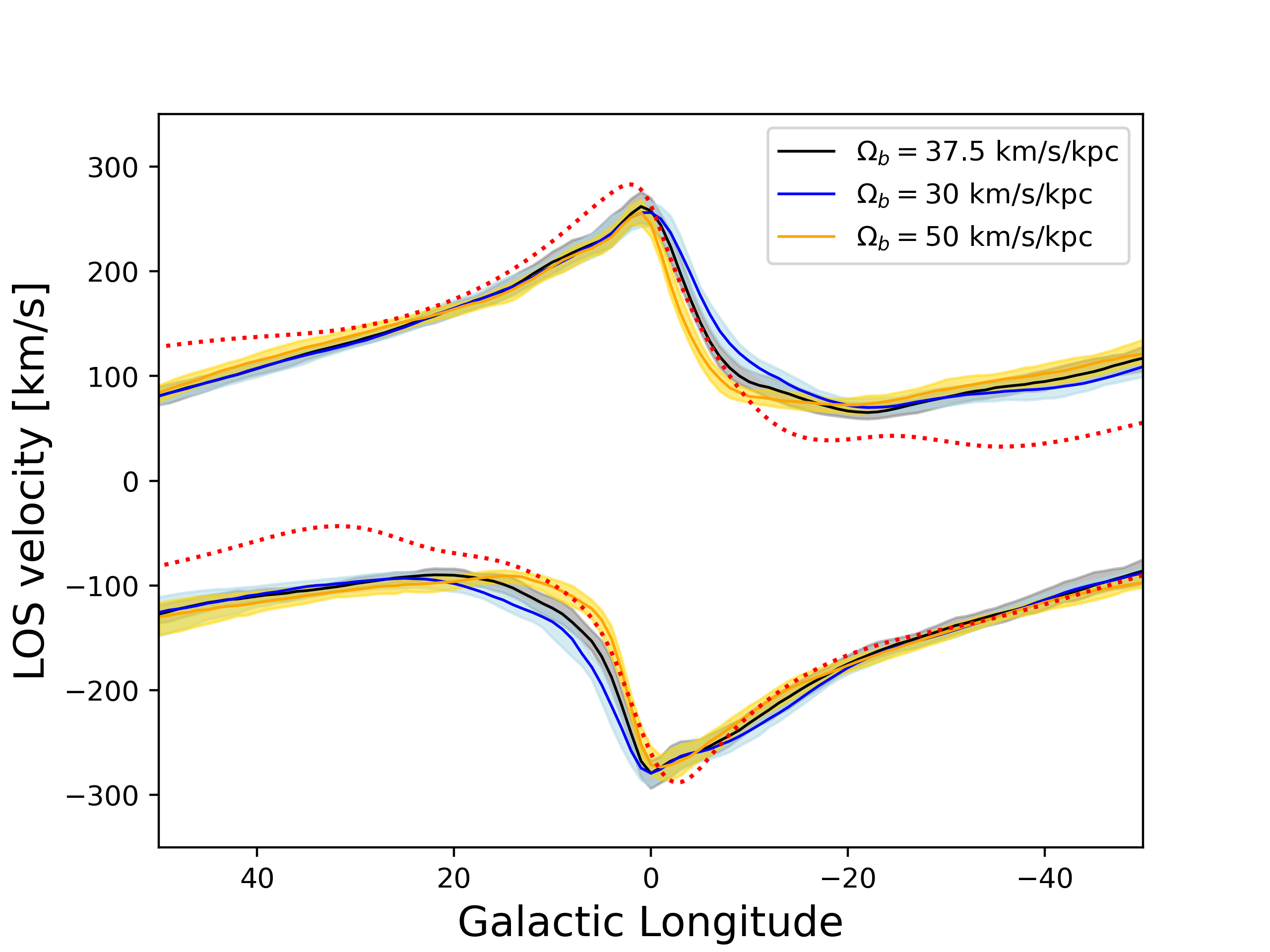}
  \includegraphics[width=0.44\textwidth]{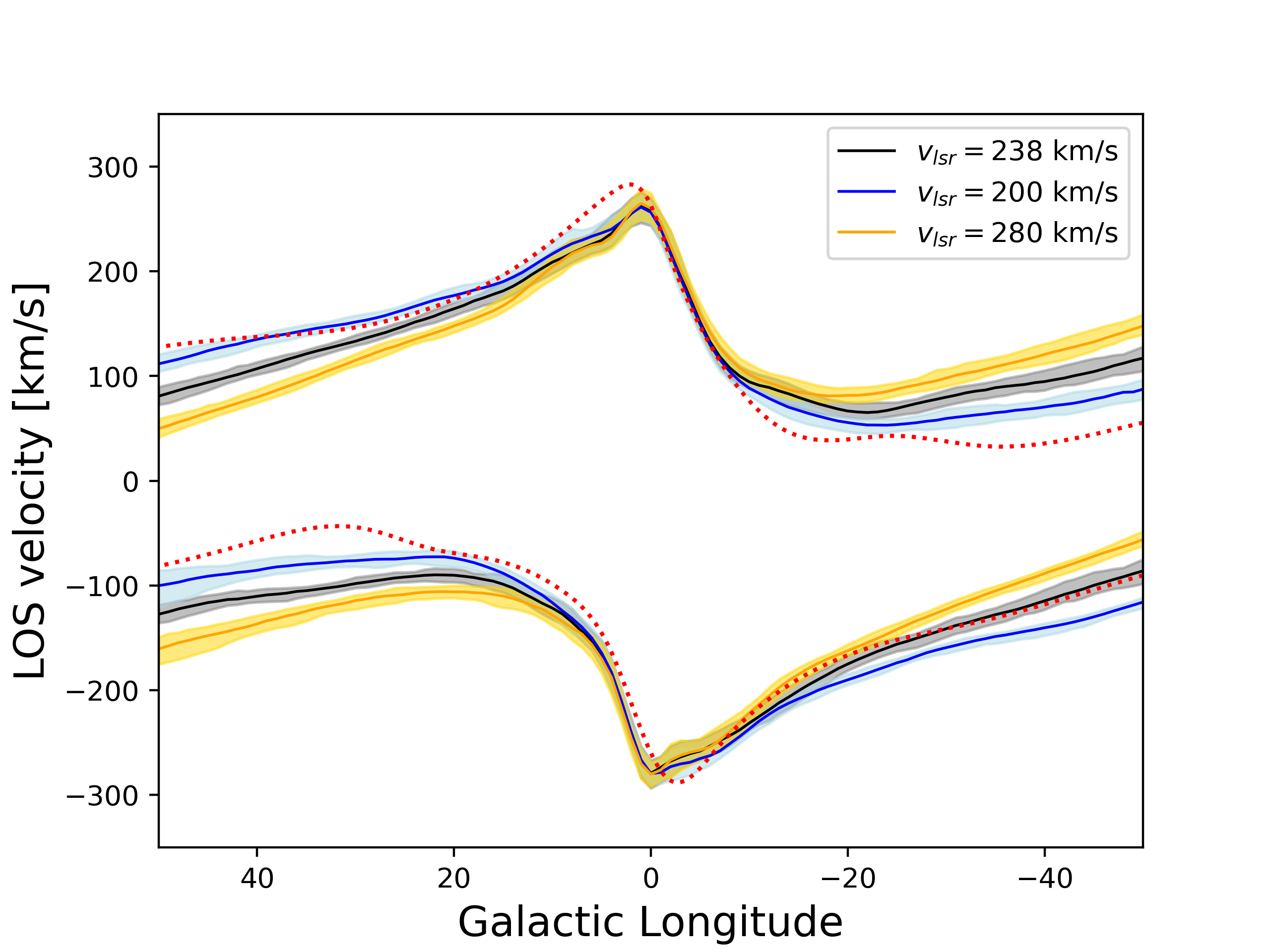}\\

  \caption{The effect of different pattern speeds $\Omega_{\rm{b}}$ and rotation velocities of the Local Standard of Rest $V_{\rm{LSR}}$ on the $\lv$ envelope. The red dashed lines in both panels represent the maser envelopes. The shaded regions represent the error bar.}
  \label{fig:omegab_vlsr}
\end{figure}

We also conduct tests on other important parameters of the model, including the pattern speed $\Omega_{\rm{b}}$ and the rotation velocity of the Local Standard of Rest $V_{\rm{LSR}}$. The results are shown in Figure~\ref{fig:omegab_vlsr}.

We find that the pattern speed only affects the envelope in the central regions of Quadrants A and C. Specifically, a lower pattern speed results in a larger envelope in these regions. {Our standard model used a value of $\Omega_{\rm{b}}=37.5\freq$}, which is consistent with previous studies \citepalias{P17}. The $\lv$ envelope generated by this value is slightly higher than that of the observational data, especially in the central region of Quadrant C, suggesting that the value may be underestimated.

As for the rotation velocity of the Local Standard of Rest, $V_{\rm{LSR}}$ affects outer regions of all quadrants of the $\lv$ envelope. However, due to the difficulty in distinguishing the best-fit value of $V_{\rm{LSR}}$, which is already well-constrained by observations, we do not consider it to be a significant factor in our analysis.

In summary, our analysis reveals that the outer regions of the $\lv$ envelope are primarily influenced by observational factors and the initial kinematic properties, while the inner regions are dominated by the impact of the monopole and quadrupole terms of the bar potential, as well as the pattern speed of the model. 

\section{Discussion}
\label{section:modification}
\subsection{Implication of the \citetalias{P17} potential}

Based on the results presented in \S~\ref{section:factor}, we can draw the conclusion that the $\lv$ envelope serves as an effective tool to constrain the Galactic bar potential. In this section, we delve into a detailed discussion of the information embedded in the $\lv$ envelope, taking the \citetalias{P17} potential as an illustrative example. We refer back to Figure~\ref{fig:compare_data}, which compares the $\lv$ envelope generated by the \citetalias{P17} potential with the observational data.

As discussed in \S~\ref{section:factor}, the outer region of $\lv$ envelope is primarily influenced by observational factors and the initial setup of the models, which are not directly associated with the Galactic bar potential. Consequently, we can ignore the discrepancies in the outer envelope where one does not expect bar orbits to reach.

We notice that the monopole term and the quadrupole length $r_q$ also have an impact on the outer region of the envelope. However, we find that a significant impact is often based on unrealistic values of these parameters. Therefore, we overlook those effects.

Regarding the inner region of the $\lv$ envelope, the main factors influencing its properties include the monopole term, the two parameters of the quadrupole term, and the bar pattern speed.


{Among the three factors, envelopes in Quadrants B and D are mainly affected by the monopole term. Therefore, the smaller envelope generated by the \citetalias{P17} potential compared to the observational data in the central region of Quadrants B suggests that the monopole (rotation curve) of the \citetalias{P17} potential may be slightly underestimated. However, we also notice that compared with the observational data, the envelope generated by the \citetalias{P17} potential is less extended in Quadrant B, but comparable in Quadrant D. This is caused by the asymmetry of the maser observation, as discussed in Appendix~\ref{section: asymmetry}.}

{On the contrary, the central envelopes of Quadrants A and C are influenced by multiple factors, including both the monopole and quadrupole terms, as well as the pattern speed. Since enlarging the monopole term, as suggested by envelopes of Quadrants B, may result in a more extended envelope in the central regions of Quadrants A and C, the existing larger envelope at these regions may indicate that the \citetalias{P17} potential tends to have a larger quadrupole strength and a possibly smaller pattern speed compared to the true values.}


As for the quadrupole length $r_q$, its response on the $\lv$ envelope is complex and non-linear, making it difficult to draw conclusions based on the envelope constraints alone.

\subsection{The \citetalias{sormani_new} potential}
\label{section:H24}

The \citetalias{P17} model included a CMC distributed as an elongated exponential disk to match the central velocity dispersion from the BRAVA data \citep{P17}. \citet{Sormani2022a} then adopted a same CMC with total mass $M_{\rm{CMC}}=2\times10^9\Msun$ and scale radius $R_{\rm{d}}=0.25\kpc$ when parametrizing the \citetalias{P17} potential. However, \citet{Li_2022} found evidence supporting an axisymmetric nuclear component in their gas model. Therefore, they replaced the original CMC in \citet{P17} with a fiducial model constrained by \citet{Sormani_2020}, which is composed of the NSC and the NSD. 

{Based on the work of \citet{Li_2022}, the \citetalias{sormani_new} potential proposed by \citet{sormani_new} combined the bar model in \citet{P17} with several other mass components, including a central black hole, an NSC, an NSD, the thin and thick stellar disks, gas disks, and a dark halo. This revised version retained the bar component of the original \citetalias{P17} potential presented in \citet{Sormani2022a}, but replaced the elongated CMC with a new set of central axisymmetric structures. }

\begin{figure*}[htbp!]
  \centering
  \includegraphics[width=0.44\textwidth]{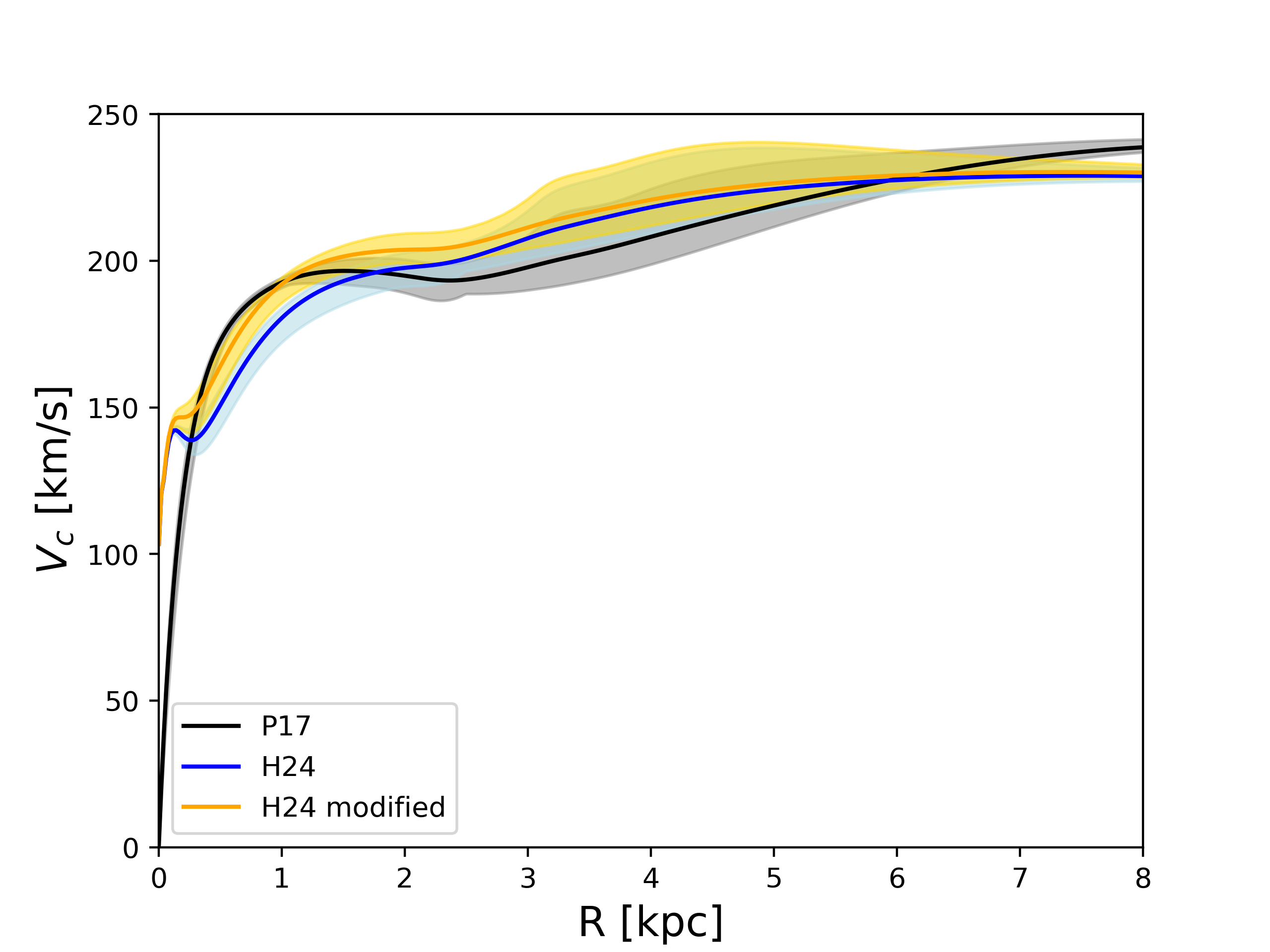}
  \includegraphics[width=0.44\textwidth]{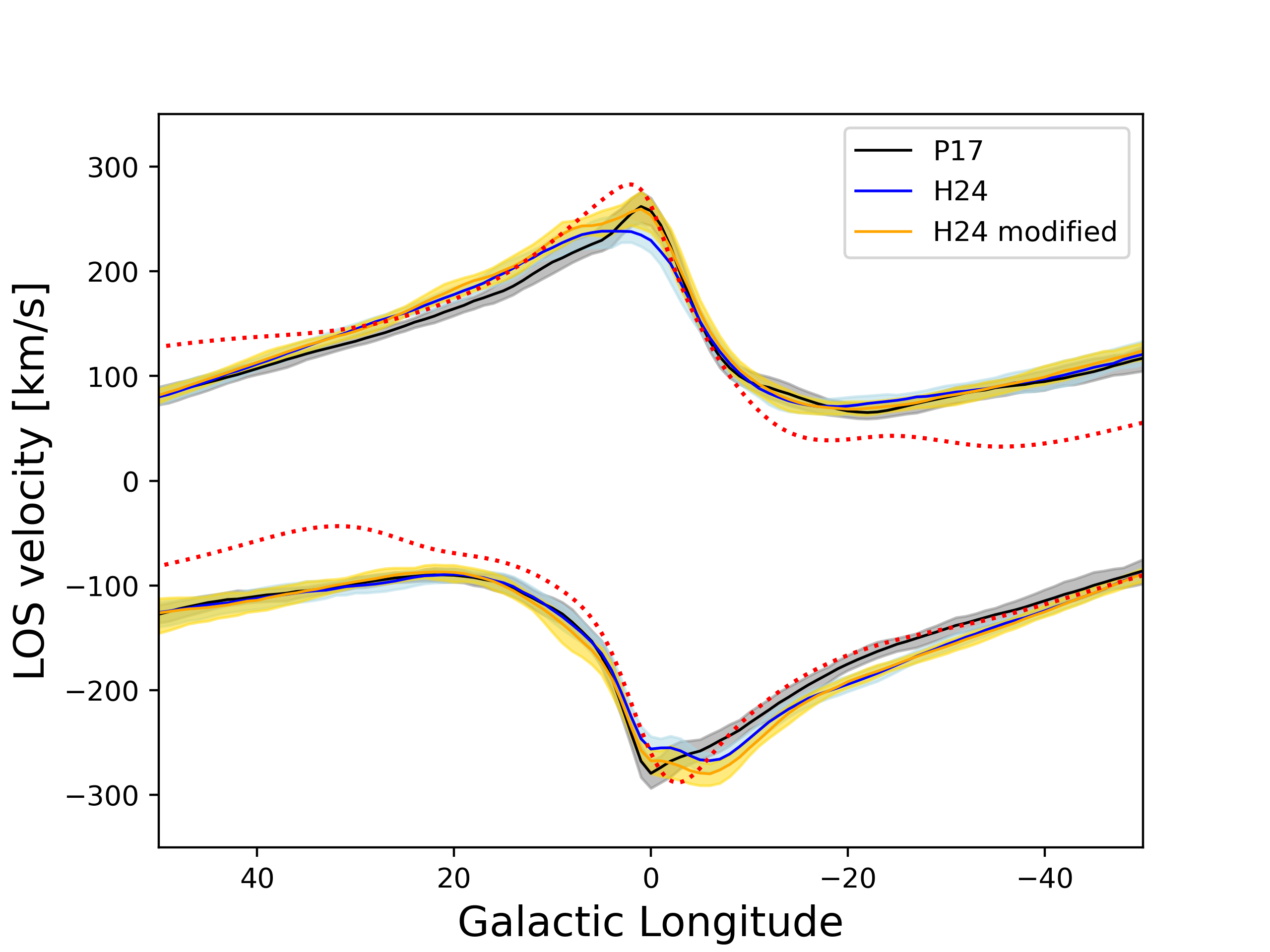}\\

  \caption{The rotation curves and the $\lv$ envelopes of the \citetalias{P17} potential, the \citetalias{sormani_new} potential and the modified \citetalias{sormani_new} potential. The red dashed line in the right panel represents the maser envelope. The shaded regions in the left panel are bounded by the maximum and minimum rotation velocity at the given radius, and the shaded regions in the right panel represent the error bar.}
  \label{fig:compare_potential}
\end{figure*}

{We compare the $\lv$ envelope generated by the \citetalias{P17} and \citetalias{sormani_new} potentials. The results are shown in Figure~\ref{fig:compare_potential}. Although the bar components in \citetalias{P17} and \citetalias{sormani_new} potentials remain unchanged, the central mass replacement affects the quadrupole term of the bar potential within $2\kpc$. However, the change is relatively small, and there is little difference of $\lv$ envelopes in Quadrants A and C. On the other hand, the rotation curve (which represents the monopole term) of the \citetalias{sormani_new} potential increases relative to that of the \citetalias{P17} potential at $2\kpc<R<5\kpc$, resulting in a larger $\lv$ envelope in Quadrants B and D. }

{Compared to the maser observation, at $10\degree<|l|<30\degree$, the new envelope generated by the \citetalias{sormani_new} potential is more extended in Quadrant D while remains comparable in Quadrant B, which is also a result of the asymmetry. Conversely, within the range of $|l|<5\degree$, the new $\lv$ envelope is less extended, particularly evident on the positive-velocity side. The prominent central mismatch could arise from a combination of various factors, including the contributions from the monopole and quadrupole terms.}

{One possible explanation for the unsatisfying fit of the \citetalias{sormani_new} potential, which results in a flat top of the $\lv$ envelope at the center, is the removal of the elongated CMC. This removal may lead to an underestimate of central bar mass of the potential, which in turn affects the overall shape of the $\lv$ envelope. We indeed see signs of a better fit if we add an additional elongated CMC to the \citetalias{sormani_new} potential, as shown in the yellow lines in Figure~\ref{fig:compare_potential}. However, we have to admit that the evidence is not particularly strong since the difference in envelopes is comparable to error bars. The details of modifying the \citetalias{sormani_new} potential by adding the additional elongated CMC are shown in Appendix~\ref{section: CMC}. }  



\subsection{The effect of the self-consistency}

{It is important to note that our test particle simulation for constructing the envelope overlooks self-consistency. We examine the impact of self-consistency on the envelope using the $N$-body simulation in \citet{shen10} at $t=800$ (hereafter S10t800).} 

\begin{figure*}[htbp!]
  \centering
  \includegraphics[width=0.88\textwidth]{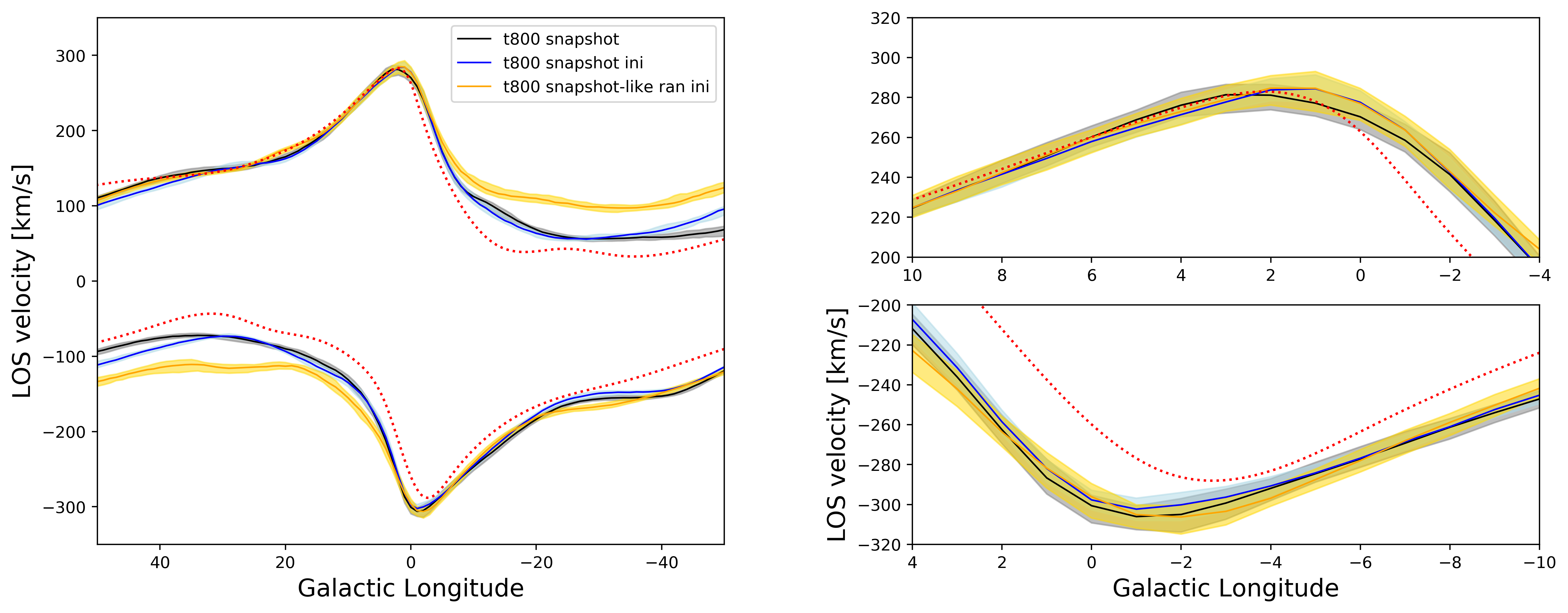}

  \caption{{The effect of self-consistencies on the $\lv$ envelope. The black lines represent the envelopes directly generated by the self-consistent \citetalias{shen10} snapshot particles. The blue envelopes are obtained by time-integrating the snapshot particles. The yellow lines show the envelopes of a self-inconsistent model constructed from integrating randomly scattered particles with snapshot-like initial conditions, i.e. the same radial and velocity dispersion profiles as the snapshot particles distribute. The left panel shows the complete $\lv$ envelopes, while the right two panels show the zoom-in envelopes in the central region. The red dashed lines represent the maser envelopes. The shaded regions represent the error bar.}}
  \label{fig:self-consistent}
\end{figure*}

{In Figure~\ref{fig:self-consistent}, we compare our envelope derived by integrating test particles in the \citetalias{shen10} potential (yellow line) with that obtained by the $N$-body particles in \citetalias{shen10} snapshot (black line). We find that if we set an initial condition with the same radial profile and the velocity dispersion profile as in the snapshot, our procedure can indeed result in a very similar central envelope shape compared with that of the snapshot. In addition, if we directly integrate the snapshot particles in the \citetalias{shen10} potential (blue line), we are also able to obtain a similar central envelope. This demonstrates that our test particle integration approach, although not fully self-consistent, can still largely reproduce the true envelope in the central bar region of the $N$-body model.}

{We notice there are also discrepancies in the outer region of the envelopes, especially in Quadrants A and C. The envelope generated by integrating randomly scattered particles appears to be more extended at $|l|>5^\circ$ in Quadrants A and C. This is because a few particles in our model may migrate from center to regions with relatively large radii. Those particles typically belong to orbits with large maximum radius (larger than $5 \kpc$) and high ellipticity values ($(R_{\rm{max}}-R_{\rm{min}})/(R_{\rm{max}}+R_{\rm{min}})>0.8$), predominantly located in Quadrants A and C with large LOS velocities after integration. Particles in the self-consistent snapshot, however, are less likely to follow such orbits as they exhibit greater stability and convergence in the phase space.}

{Since the \citetalias{P17} snapshot is temporarily inaccessible, we can only minimize the difference between our model and the snapshot by adopting similar radial and velocity dispersion profiles. The radial profile can be easily derived from the surface density of the P17 potential, which is available in the {\tt Agama} package. The velocity dispersion profile, however, is not easy to approximate. In this paper, we use the Gaia DR3 data to mimic the snapshot velocity dispersion profile. We admit that the difference between envelopes of our model and the P17 snapshot exist inevitably, but those uncertainties should be within error bars in the central bar region.}

\section{Conclusion}
\label{section:conclusion}

In this study, we propose a novel method for constraining the Milky Way bar potential using the $\lv$ envelope of SiO maser stars. Instead of using only one type of orbit in a given gravitational potential to trace the high-velocity stars in the $\lv$ diagram as \citet{Habing2016} did, we demonstrate that the $\lv$ envelope can capture comprehensive information about all orbital families within the bar potential. Consequently, it serves as a valuable tool for assessing the overall properties of the bar potential.

Using preliminary observational data from the Bulge Asymmetries and Dynamical Evolution (BAaDE) survey, which includes 15,207 SiO maser stars, we show that these stars exhibit a relatively well-defined envelope in the $\lv$ diagram.

We conduct a comparison between the observed $\lv$ envelope and that generated by a widely-used \citetalias{P17} potential, utilizing the same sample size. We establish initial conditions of orbits based on the velocity dispersion map of the Gaia DR3 data provided by \citet{gaiadr3}. Overall, the $\lv$ envelope generated by the model aligns well with the observational results, indicating a good agreement between \citetalias{P17} potential and the observations. 

However, small inconsistencies exist. We demonstrate that the outer discrepancies are attributed to observational factors and the initial kinematic properties of the test particles, while the inner discrepancies are mainly due to factors related to the bar potential, such as the monopole and quadrupole terms of the bar potential, as well as the pattern speed of the model.  

Our exploration based on the comparison between the $\lv$ envelope of the model and the data suggests that the \citetalias{P17} potential yields a smaller rotation curve and a slightly larger quadrupole strength relative to the actual Milky Way potential, and that the pattern speed we set in the model may be underestimated.

For further application of the $\lv$ envelope, we adopted the \citetalias{sormani_new} potential in our analysis, which is an modification of the \citetalias{P17} potential. However, despite the optimization of the CMC in the \citetalias{sormani_new} potential, an excessive amount of central mass has been eliminated, resulting in a poor fit to the central $\lv$ envelope. {We see signs that reintroducing an additional CMC can help alleviate the issue. Nevertheless, it must be acknowledged that the evidence is modest, as the distinctions in envelopes are comparable to the error bar.}


In summary, the $\lv$ envelope serves as a powerful tool to probe the Milky Way potential and its dynamical structure. However, some limitations still exist in the method. Firstly, the choice of initial conditions greatly influences the resulting envelope, emphasizing the need for accurate estimation of the velocity dispersion distribution prior to fitting. {Secondly, certain detailed parameters of components within the Galactic potential cannot be fully constrained by the $\lv$ envelope alone.} Therefore, additional tools are required. One potential approach is to consider the central distribution of stars in the $\lv$ diagram. Furthermore, the presence of asymmetry in the observational envelope significantly impacts the quality of the fitting and may warrant further investigation and detection.

\begin{acknowledgements}
This paper uses data products obtained with instruments run by the National Radio Astronomy
Observatory (NRAO). The National Radio Astronomy Observatory is a facility of the National Science Foundation operated under cooperative agreement by Associated Universities, Inc.
The research presented here is partially supported by the National Key R\&D Program of China under grant No. 2018YFA0404501; by the National Natural Science Foundation of China under grant Nos. 12103032, 12025302, 11773052, 11761131016; by the ``111'' Project of the Ministry of Education of China under grant No. B20019; and by the Chinese Space Station Telescope project. J.S. acknowledges support from a {\it Newton Advanced Fellowship} awarded by the Royal Society and the Newton Fund. This work made use of the Gravity Supercomputer at the Department of Astronomy, Shanghai Jiao Tong University, and the facilities of the Center for High Performance Computing at Shanghai Astronomical Observatory. 
\end{acknowledgements}

\newpage

\appendix
 \renewcommand{\theequation}{\thesection.\arabic{equation}}





\section{The asymmetry of the envelope}
\label{section: asymmetry}

We note that the \citetalias{P17} potential model generates a smaller envelope in Quadrant B, but a comparable one in Quadrant D compared with the observational data, as shown in Figure~\ref{fig:compare_data}. These discrepancies can be related to the different asymmetry behaviour between the $\lv$ envelopes generated by the model and the observational data.

Regarding the envelope of the model, we find that its central asymmetry is only caused by the tangential motion of LSR with a velocity of $U_\odot=10.6\kms$. This tangential motion causes a downward shift of the entire envelope in the $\lv$ plane, approximately by $10\kms$. 

However, the observational maser envelope exhibits a different asymmetry behavior. When applying a point symmetry transformation, the positive-longitude side of the envelope appears to have a larger extent compared to the negative-longitude side.

It is worth noticing that the gas envelopes shown in the lower panel of Figure~\ref{fig:maser_lv} also exhibit asymmetry. This asymmetry is caused by the asymmetrical nature of the spirals, which manifest as tangent points in Quadrants B and D of the $\lv$ envelope for different spirals. However, since maser stars are in the very late stages of Asymptotic Giant Branch (AGB), they are more evolved stars and may not be able to trace spirals due to dynamic relaxation over their lifetimes. 

The origin of the asymmetry of the maser envelope is still under discussion. {It could be related to the larger sample size at $20\degree<l<40\degree$ than that at $-40\degree<l<-20\degree$, which is possibly resulted from the sampling of the deeper MSX survey regions \citep{maser_distance_2024}. While we have verified that the number of particles in our model does not affect the envelope, the different scattering of the maser samples in the velocity field at the positive-longitude regions compared to the negative-longitude regions could still cause the asymmetry.} 

In addition, observational uncertainty of high-velocity stars could also be taken into consideration. The poor match of the $\lv$ envelope due to the asymmetry, especially in Quadrant B, is worth more studies once the velocities for $l>40$ are re-confirmed in future work. 
\section{the $\lv$ envelope of an axisymmetric disk potential}
\label{appendix: disk}

We also test the $\lv$ envelope of an axisymmetric disk potential. The axisymmetric disk potential is simply set by removing the bar component of the \citetalias{P17} potential.

\begin{figure}[htbp!]
  \centering
  \includegraphics[width=0.44\textwidth]{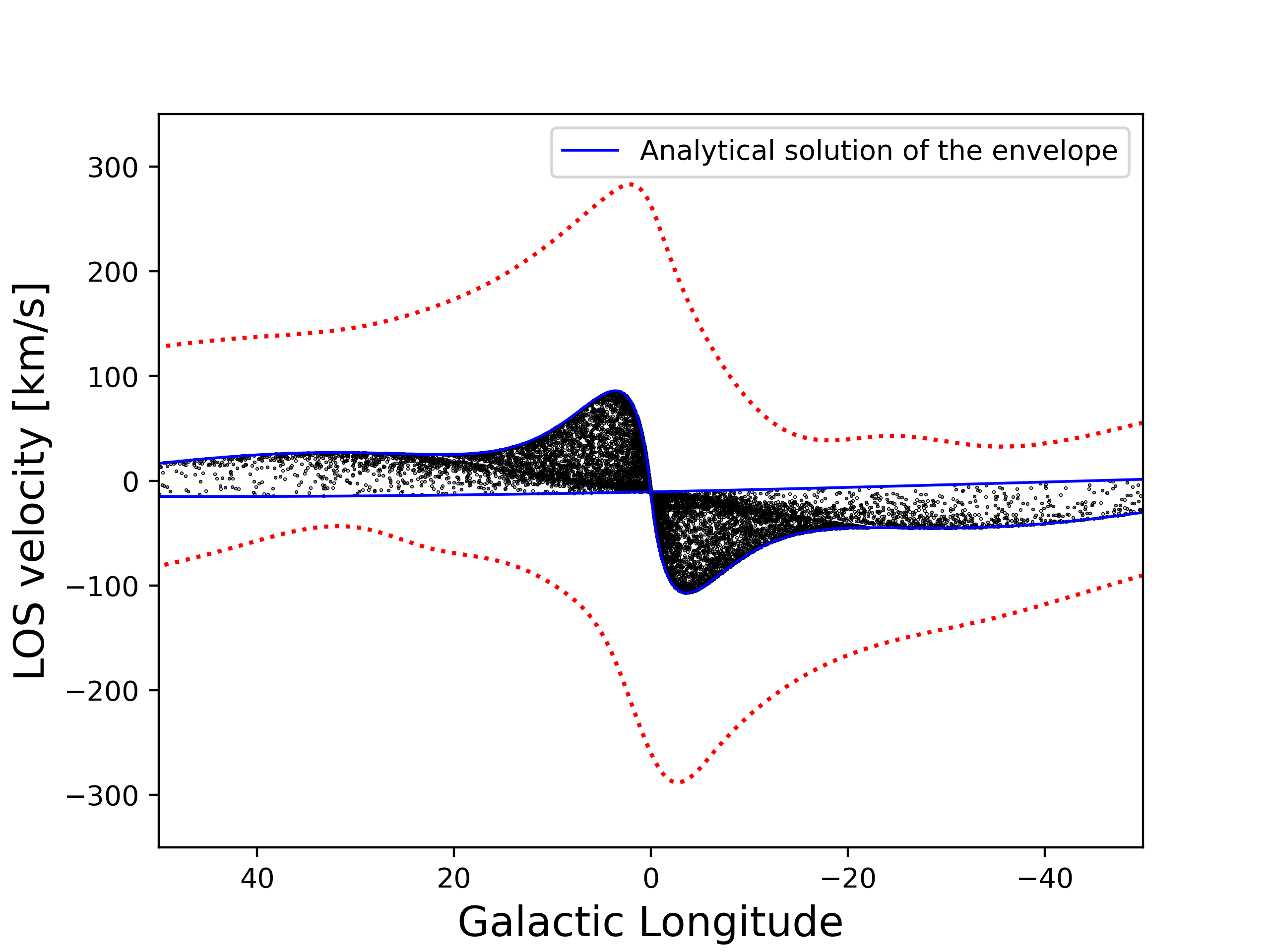}
  \includegraphics[width=0.44\textwidth]{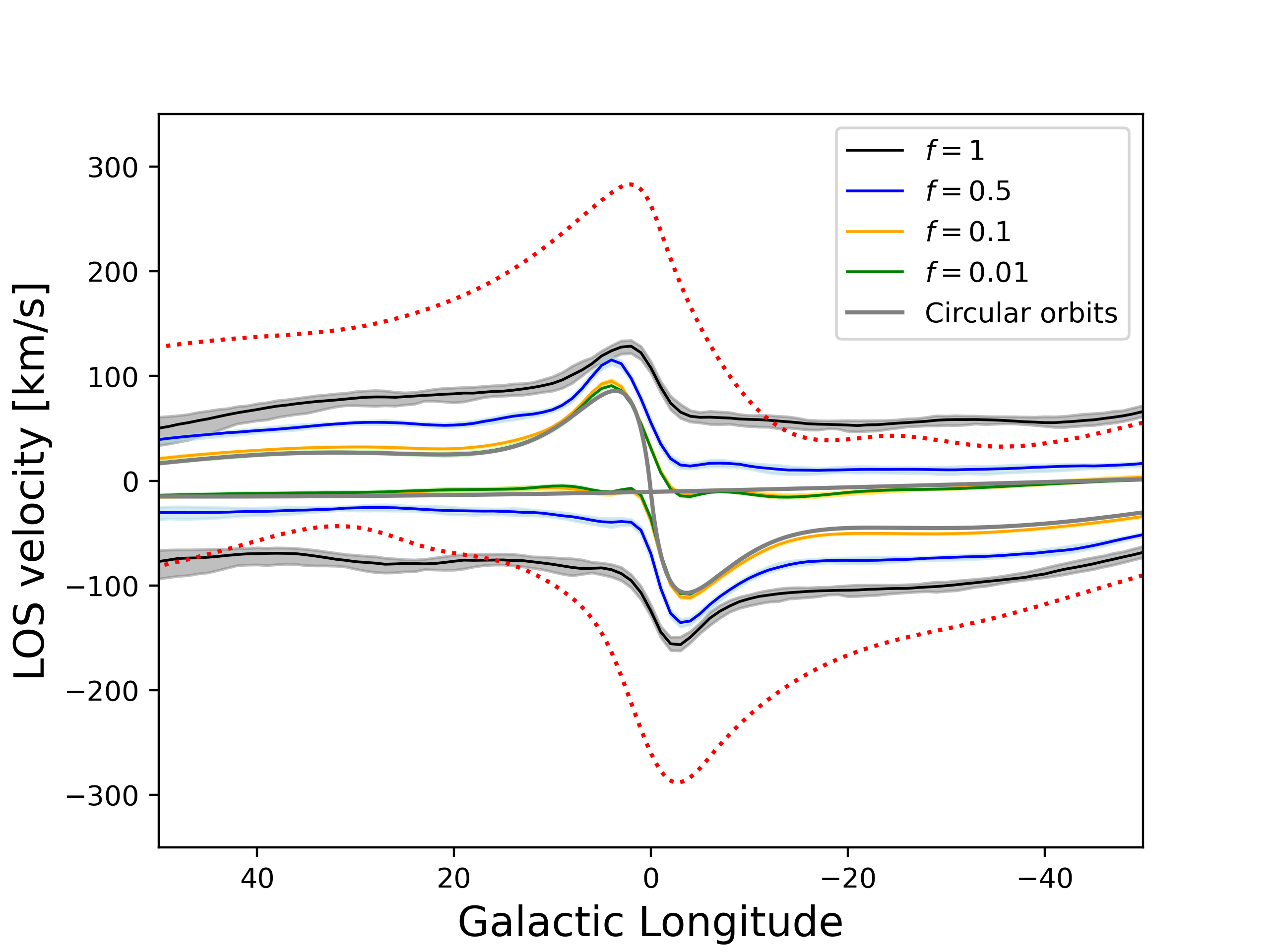}\\
  \caption{The $\lv$ envelope generated by an axisymmetric disk potential. The left panel shows the envelope of circular orbits only. The black points are test particles belonging to the circular orbits integrated from the initial conditions with an exponential distribution of radius. The blue line is the analytical result from Equation~\ref{eq:circular_vl}. The right panel shows envelopes of models with different initial velocity dispersions, where both circular and non-circular orbits are included. $f=1$, $0.5$, $0.1$, and $0.01$ indicate that models have the initial velocity dispersions multiplying the standard version by $1$, $0.5$, $0.1$, and $0.01$, respectively. The grey line is the analytical maximum velocity for circular orbits only. The shaded regions represent the error bar. The red dashed lines in both panels represent the maser envelope.}
  \label{fig:analy_disk_rc}
\end{figure}

If only considering circular orbits, since no bar exists in the disk potential, we can use Equation~\ref{eq:circular_vl} to obtain the analytical solution of the $\lv$ envelope, as shown in the left panel of Figure~\ref{fig:analy_disk_rc}. The analytical solution of the envelope well covers all the test particles of the circular orbits. Note that the LOS velocity of the envelope is not zero in Quadrant C. This is due to the solar motions relative to the Local Standard of Rest, which make the overall envelope move slightly downwards.

Then we consider the non-circular orbits. The right panel in Figure~\ref{fig:analy_disk_rc} shows the envelope comparison between different settings of initial velocity dispersion. As the initial velocity becomes smaller, the envelope will be closer to the analytical solution of circular orbits. This is consistent to the results in \S~\ref{section:factor} that circular orbits tend to lie in Quadrants B and D of the $\lv$ diagram. 

\section{The central mass component}
\label{section: CMC}

\begin{table}[htbp!]
\centering
\caption{Chi-square values of the \citetalias{sormani_new} potential and nine modification potentials. }
\begin{tabular}{@{}|c|c|c|c|@{}}
\hline
Models & $M_{\rm{CMC}}$ ($\times10^9\Msun$) & $R_{\rm{d}}$ ($\kpc$) & $\chi^2$ \\ \hline
\citetalias{sormani_new}     & 0                          &                       & 213       \\ \hline
m1      & 0.5                         & 0.15                  & 177      \\ \hline
m2      & 0.5                         & 0.25                  & 190      \\ \hline
m3      & 0.5                         & 0.45                  & 160      \\ \hline
m4      & 1                           & 0.15                  & 152      \\ \hline
m5      & 1                           & 0.25                  & 163      \\ \hline
m6      & 1                           & 0.45                  & 146      \\ \hline
m7      & 1.5                         & 0.15                  & 203     \\ \hline
m8      & 1.5                         & 0.25                  & 207      \\ \hline
m9      & 1.5                          & 0.45                  & 223     \\ \hline
\end{tabular}
\label{table:CMC_model}
\end{table}

We add back an elongated CMC to the \citetalias{sormani_new} potential to minimize the envelope difference compared to the maser observation.
We vary $M_{\rm{CMC}}$ to be $0.5\times10^9,1\times10^9,1.5\times10^9\Msun$, as well as $R_{\rm{d}}$ to be $0.15$, $0.25$, $0.45\kpc$, as shown in Table~\ref{table:CMC_model}. This leads to nine different models. To determine the best-fit model, we compare the $\lv$ envelopes of the nine models with the maser data and employ a similarity measurement. By evaluating the degree of similarity between the model predictions and the observed data, we can identify the model that provides the overall best fit.

\begin{figure*}[htbp!]
  \centering
  \includegraphics[width=0.88\textwidth]{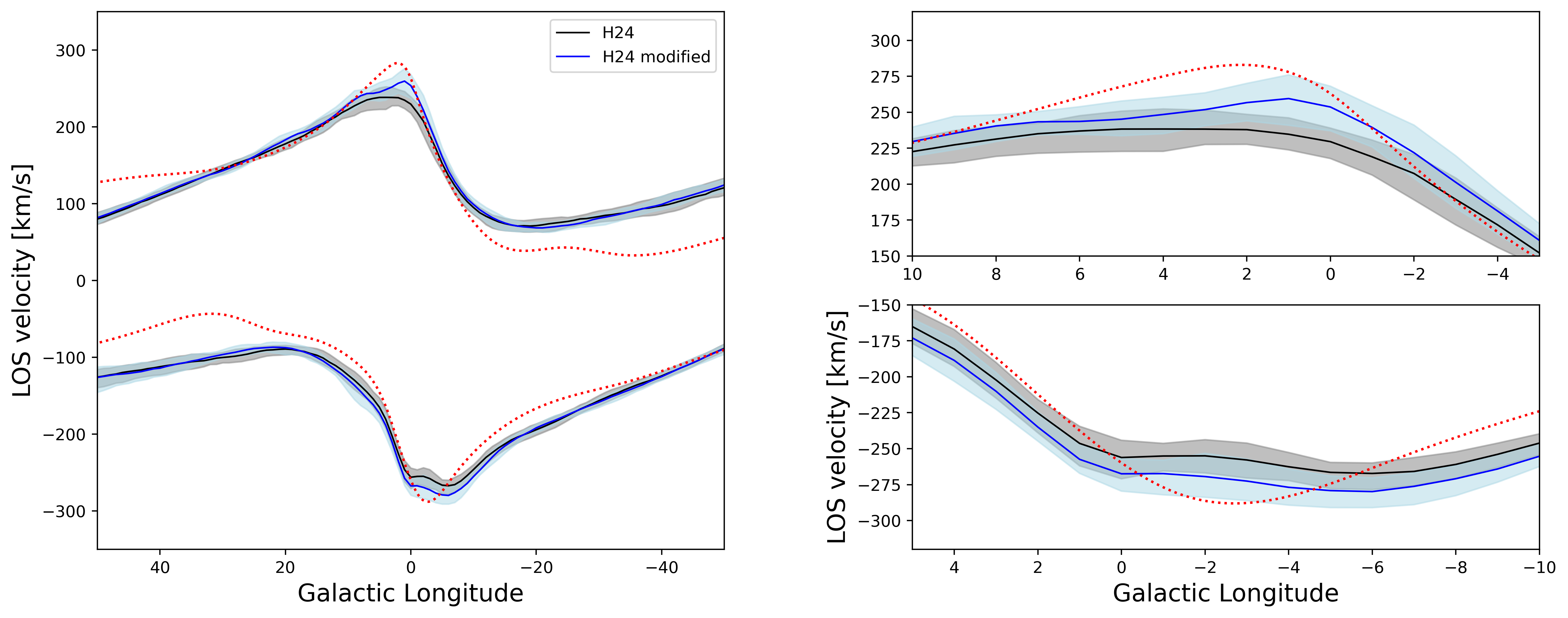}

  \caption{{The $\lv$ envelopes generated by the \citetalias{sormani_new} potential and the best-fit model. The left panel shows the complete $\lv$ envelopes, while the right two panels show the zoom-in envelopes in the central region ($-5^\circ<l<10^\circ$ for positive LOS velocities and $-10^\circ<l<5^\circ$ for negative LOS velocities). The red dashed lines represent the maser envelopes. The shaded regions represent the error bar.}}
  \label{fig:best-fit}
\end{figure*}

{For the similarity measurement, since the distribution of CMC mainly affects the bar region, we focus on the very central envelope fit ($-5^\circ<l<10^\circ$ for positive LOS velocities and $-10^\circ<l<5^\circ$ for negative LOS velocities). We calculate the chi-square values of the nine models and the one generated by the \citetalias{sormani_new} potential. Based on the analysis, we identify the best-fit model with the smallest chi-square value and present its $\lv$ envelope in Figure~\ref{fig:best-fit}.}

{We find that the envelope constraint indicates an additional CMC with $M_{\rm{CMC}}=1\times10^9\Msun$ and $R_{\rm{d}}=0.45\kpc$. However, we admit that the variation of the chi-square values in different models is modest. If considering other systematic effect such as the setting of initial conditions and the self-consistency, the evidence of a better fit envelope is not strong, as acknowledged in \S~\ref{section:H24}.} 





\bibliographystyle{aasjournal}


\end{document}